\documentclass[mnsc,nonblindrev]{informs4rad}

\newif\ifblind
\blindfalse  % switch to \blindtrue for the anonymous version

\newcommand{\revcolor}[1]{{\color{black}#1}}

\OneAndAHalfSpacedXI

\usepackage[showdeletions]{color-edits}

\addauthor{FE}{blue}
\makeatletter
\renewcommand{\FEdelete}[1]{}
\renewcommand{\FEedit}[1]{#1}
\makeatother

\usepackage{booktabs}
\usepackage{mathtools}

\usepackage{xcolor}
\usepackage{natbib}
\usepackage{mathrsfs}
\usepackage{dsfont}
\definecolor{cornellred}{rgb}{0.7, 0.11, 0.11}
\definecolor{maroon}{rgb}{0.52, 0, 0}
\definecolor{dgreen}{rgb}{0.0, 0.5, 0.0}
\definecolor{ballblue}{rgb}{0.13, 0.67, 0.8}
\definecolor{royalblue(web)}{rgb}{0.25, 0.41, 0.88}
\definecolor{bleudefrance}{rgb}{0.19, 0.55, 0.91}
\definecolor{royalazure}{rgb}{0.0, 0.22, 0.66}
\usepackage{hyperref}
\hypersetup{
	colorlinks = true,
	linkcolor=maroon,
	urlcolor=royalazure,
	citecolor=royalazure,
	linkbordercolor = {maroon}
}
\usepackage{cleveref}
\usepackage{float}
\usepackage{enumitem}
\usepackage{xcolor}
\usepackage{booktabs} % For formal tables
\usepackage{amsmath}
\usepackage{subcaption}
\usepackage{tikz}
\usetikzlibrary{positioning}
\usetikzlibrary{calc}

\usepackage{csquotes}
\makeatletter
\renewenvironment*{displayquote}
  {\begingroup\setlength{\leftmargini}{0.7cm}\csq@getcargs{\csq@bdquote{}{}}}
  {\csq@edquote\endgroup}
\makeatother

\renewcommand{\qed}{$\hfill\square$}

\newenvironment{proofof}[1]{%
  \Trivlist
  \item[\hskip\labelsep {\it #1.}]\ignorespaces
}{\hfill \qed
\endTrivlist
\addvspace{0pt}
}
\usepackage[ruled,vlined,linesnumbered]{algorithm2e}

\SetCommentSty{mycommfont}

\DontPrintSemicolon
\SetAlgoVlined
\SetAlFnt{\small}
\SetAlCapFnt{\small}
\SetAlCapNameFnt{\small}
\SetAlCapHSkip{0pt}
\IncMargin{-\parindent}
\newcommand{\prob}{\text{I\kern-0.15em P}}
\setcitestyle{authoryear}

\allowdisplaybreaks
\newtheorem{theorem}{Theorem}[section]
\usepackage{thm-restate}
\newtheorem{lemma}[theorem]{Lemma}
\newtheorem{proposition}[theorem]{Proposition}
\newtheorem{corollary}[theorem]{Corollary}

\newtheorem{definition}{Definition}[section]

\newtheorem{remark}[definition]{Remark}
\newtheorem{example}[definition]{Example}
	\newtheorem{assumption}[definition]{Assumption}

\newcommand{\E}{\mathbb{E}}
\newcommand{\Prob}{\mathbb{P}} % (requested) use \Prob instead of \P
\newcommand{\Ind}[1]{\mathbb{I}\{#1\}}

\newcommand{\NumRiders}{m}
\newcommand{\NumDrivers}{n}
\newcommand{\RiderSet}{\mathcal{R}}
\newcommand{\DriverSet}{\mathcal{D}}

\newcommand{\wSym}{w} % weight symbol
\newcommand{\pSym}{p} % acceptance-probability symbol
\newcommand{\Weight}[2]{\wSym_{#1,#2}} % w_{i,j}
\newcommand{\AccProb}[2]{\pSym_{#1,#2}} % p_{i,j}

\newcommand{\wmax}{1} % w.l.o.g. weights in [0,1]

\newcommand{\wSingle}[1]{\wSym_{#1}} % w_j
\newcommand{\pSingle}[1]{\pSym_{#1}} % p_j

\newcommand{\AccRV}[2]{X_{#1,#2}} % acceptance indicator X_{i,j}
\newcommand{\Bern}{\mathrm{Bern}}

\newcommand{\Val}{F}

\newcommand{\Fi}[2]{\Val_{#1}\!\left(#2\right)}

\newcommand{\Tau}{\tau}
\newcommand{\TauOf}[1]{\Tau\!\left(#1\right)}

\newcommand{\del}{\delta}      % (typically multiplicative slack)

\newcommand{\tauTilde}{\tilde{\tau}}
\newcommand{\HighSet}{T^{\mathrm{H}}}
\newcommand{\MidSet}{T^{\mathrm{M}}}
\newcommand{\LowSet}{T^{\mathrm{L}}}
\newcommand{\HatL}{\hat{L}}
\newcommand{\MidSetBucket}[1]{T^{\mathrm{M},#1}}
\newcommand{\SMid}{S^{\mathrm{M}}}

\newcommand{\xhdr}[1]{\noindent\textbf{#1}}

\MANUSCRIPTNO{}
\begin{document}
\RUNTITLE{Non-Exclusive Notifications for Ride-Hailing at Lyft I: Single-Cycle Approximation Algorithms}
\TITLE{Non-Exclusive Notifications for Ride-Hailing at Lyft I: Single-Cycle Approximation Algorithms}
\RUNAUTHOR{Ekbatani et al.}

\ARTICLEAUTHORS{

\AUTHOR{Farbod Ekbatani, Rad Niazadeh}
\AFF{University of Chicago Booth School of Business, Chicago, IL\\
\EMAIL{fekbatan@chicagobooth.edu, rad.niazadeh@chicagobooth.edu}}

\AUTHOR{Shruti Penumatsa}
\AFF{University of California Berkeley, Berkeley, CA\\
\EMAIL{shruti\_penumatsa@berkeley.edu}}

\AUTHOR{Mehdi Golari, Romain Camilleri, Titouan Jehl, Chris Sholley, Matthew Leventi, Theresa Calderon}
\AFF{Lyft, Inc.\\ \EGT\EMAIL{mehdig@lyft.com, rcamilleri@lyft.com, titouanj@lyft.com, chris@lyft.com, mleventi@lyft.com, tcalderon@lyft.com}}

\AUTHOR{Angela Lam, Paul Havard Duclos, Tim Holland, James Koch, Shreya Reddy}
\AFF{Lyft, Inc.\\ \EGT\EMAIL{alam@lyft.com, paul.havardduclos@gmail.com,tholland@lyft.com, jkoch@lyft.com, sreddy@lyft.com}}
}

\ABSTRACT{
Ride-hailing platforms increasingly rely on non-exclusive notifications—broadcasting a single request to multiple drivers simultaneously—to mitigate inefficiencies caused by uncertain driver acceptance. In this paper, the first in a two-part collaboration with Lyft, we formally model the \emph{Notification Set Selection Problem} for a single decision cycle, where the platform determines the optimal subset of drivers to notify for each incoming ride request. We analyze this combinatorial optimization problem under two contention-resolution protocols: \emph{First Acceptance (FA)}, which prioritizes speed by assigning the ride to the first responder, and \emph{Best Acceptance (BA)}, which prioritizes match quality by selecting the highest-valued accepting driver.

We show that welfare maximization under both mechanisms is \emph{NP-hard}, ruling out a polynomial-time exact algorithm unless $\mathrm{P}=\mathrm{NP}$. Despite this hardness, we establish several positive algorithmic results. For FA, we develop a Polynomial Time Approximation Scheme (PTAS) for the single-rider problem. For the general matching setting, we provide a $(1-1/e)$-approximation algorithm that applies to both FA and BA. Along the way, we show that the FA valuation function constitutes a novel discrete choice model with structural properties of independent interest. For BA, we further prove that the welfare objective is monotone and submodular and exploit additional structure to obtain an approximation guarantee that surpasses the classical $(1-1/e)$ barrier. Finally, when acceptance probabilities are independent of the drivers, we show that the BA problem can be solved exactly in polynomial time through a linear programming formulation. We evaluate our algorithms on synthetic instances and on instances calibrated using real ride-sharing data from Lyft, demonstrating strong empirical performance.
}
\maketitle

\newpage

\section{Introduction}
\label{sec:intro}
Modern ride-hailing platforms rely on real-time matching algorithms to pair riders with nearby drivers. Matching decisions are typically executed in discrete cycles, where the platform observes a batch of active ride requests and a pool of available drivers, computes a dispatch decision, and issues notifications. Traditionally, these systems rely on one-to-one matching protocols, also known as \emph{exclusive dispatch (ED)}: each ride is
offered to a single driver at a time. If the driver accepts, the match is finalized; if they reject or fail to respond, the platform must proceed sequentially by recomputing a new match option in future cycles (when the ride request becomes active again for matching) and offering the ride to this ``next best candidate'' driver.

While ED is operationally simple and minimizes the coordination overhead, it faces a growing challenge in this gig economy: stochastic acceptance behavior. Drivers are gig-economy workers who retain discretion over which trips to accept. As a result, rejections, ignored offers, and timeouts are common in practice and can arise from fatigue, destination preferences, traffic conditions, or other considerations. As rejection rates increase, the sequential nature of \FEedit{exclusive dispatch} creates inefficiencies: riders experience prolonged waiting times while the system cycles through unresponsive drivers, leading to higher cancellation rates from riders and platform friction (e.g., additional computation or increasing congestion in the matching pipeline).

To hedge against uncertain driver \FEdelete{compliance}\FEedit{acceptance}, platforms increasingly employ \emph{non-exclusive dispatch (\FEedit{NED}) notifications}: instead of offering a ride request to a single driver, the platform broadcasts the same request to a \emph{set} of candidate drivers simultaneously.  The
operational rationale of this one-to-many matching approach is straightforward: parallelizing invitations increases the probability that at
least one driver accepts quickly, reducing match latency and improving reliability.  At the same
time, \FEedit{NED} notifications introduce a new complex combinatorial challenge: the \emph{Notification Set Selection} problem. Given a single decision cycle and a set of ride requests and available drivers with heterogeneous acceptance rates, which subset of drivers should the platform notify for each ride? The platform also faces a coordination challenge: the \emph{Contention Resolution} protocol. Given that multiple drivers may accept the same request, what protocol should the platform employ to determine which accepting driver receives the trip and when the decision is finalized?

In this paper---as part of our research collaboration with Lyft---we develop a framework to formalize and study the above algorithmic questions.~\footnote{In the current paper, we isolate the single-cycle problem that sits at the core of any NED pipeline. In a more applied companion paper of this work (Part II; \ifblind\cite{LyftII2026companion_Blind}\else\cite{LyftII2026companion_NonBlind}\fi) we study NED in a dynamic marketplace using simulations and a stylized macro-model, focusing on long-run
effects and implementation trade-offs. In particular, Part~II highlights that while broadcasting can temporarily reserve multiple drivers for the same request---which may thin effective supply and shape long-run performance---still NED improves throughput, match quality and match time over ED.}
We assume that acceptance decisions are stochastic with heterogeneous probabilities that depend on the rider-driver pair and independent across drivers, and that each driver can be notified at most once per cycle. The goal of the platform is to choose disjoint notification sets that maximizes the expected total \emph{welfare} in that cycle. Welfare is defined based on the \emph{matching scores} between riders and drivers (which depend on the distance, price, and other features of the rider-driver pair) and is the sum of the scores of the riders-diver pairs that are eventually matched through notifications sent in that cycle. Unlike ED, this objective is no longer linear in the selected rider--driver pairs; instead, it depends on the entire notification set and the chosen contention resolution rule.

We analyze the above combinatorial optimization problem under two distinct contention-resolution protocols. These two protocols, defined below, reflect the fundamental trade-off between match speed and match quality and are common practices in the industry:
\begin{itemize}[align=left, labelwidth=1em,
  labelsep=0.1em,
  leftmargin=1em,
  itemsep=0.1em]
    \item \textit{First Acceptance (FA):} The platform assigns the ride to the first driver who responds. This models a fast dynamic, prioritizing speed but potentially sacrificing match quality (e.g., a closer driver might be slower to interact with the app than a distant one).
    \item \textit{Best Acceptance (BA):} The platform collects all responses and assigns the ride to the highest-scoring driver among those who accepted. This prioritizes quality but introduces a mandatory waiting period.
\end{itemize}
These two mechanisms induce fundamentally different objective functions over notification sets. Under BA, adding more drivers can only improve the set’s value (the score of the selected driver in expectation), and the resulting valuations admit strong diminishing-returns structure (also known as \emph{submodularity}). Under FA, the valuation can be \emph{non-monotone}: notifying an additional low-value but high-acceptance driver can reduce expected welfare by increasing the probability that a higher-value driver loses contention. Also, such valuations can be \emph{non-submodular}. These
non-monotonicity and non-submodularity properties of valuations under FA are the main algorithmic challenges in the paper and central to our results. We also note that this new algorithmic landscape is in stark contrast to the traditional  ED, which can be cast as the maximum edge-weighted matching problem and is polynomial-time solvable. Given these challenges, we ask the following research questions:
\begin{displayquote}
\emph{Are \FEedit{NED} notification welfare maximization problems under FA and BA computationally hard to be solved exactly (or approximately)---in contrast to \FEedit{ED} which is polynomial-time solvable? If yes, can we design polynomial-time approximation algorithms for these problems?} 
\end{displayquote}

\subsection{Our Contributions}
We provide a comprehensive theoretical analysis of the \FEedit{NED} notifications welfare maximization problem defined above. Our main contributions map the computational landscape of this problem under both FA and BA protocols for resolving the contention:

\smallskip
\xhdr{Hardness of welfare maximization.} We prove that welfare maximization is \emph{NP-hard} under either contention protocol
(\Cref{thm:hardness}), even under highly restricted instances: there are only two riders, the riders are identical, all edge weights are unit, and the acceptance probabilities depend only on the driver. In particular, this rules out a polynomial-time exact algorithm unless $\textsf{P}=\textsf{NP}$, and justifies the focus on designing constant-factor approximation algorithms.

\smallskip
\xhdr{Single Rider: a PTAS for FA via threshold structure.} One of the main technical challenge of the paper arises under FA, where the valuation is neither monotone
nor submodular.  Even for a single rider, the platform must decide \emph{which} drivers to
include, balancing high match value against the risk of dilution under FA, and this problem is NP-hard (\Cref{thm:NPhard}). Our first main algorithmic result is a \emph{Polynomial Time Approximation Scheme (PTAS)} for the single-rider FA problem (\Cref{sec:single-ride}).

The key insight behind our PTAS is a structural characterization of optimal solutions: there exists an endogenous threshold $\tau^*$ such that
(i) every driver with weight strictly larger than $\tau^*$ must be included,
(ii) no driver with weight smaller than roughly $\tau^*/3$ is included, and
(iii) only a ``medium'' weight band around $\tau^*$ requires careful selection.
We discretize this medium band into $O(1/\delta)$ buckets, guess the number of selected drivers
per bucket, and show that within each bucket it is optimal (after rounding) to keep the drivers with
largest acceptance probabilities.  This approach yields a $(1-\delta)$-approximation in time $n^{O(1/\delta)}$, where $n$ is the number of drivers (Algorithm~\ref{alg:single} and \Cref{thm:alg1}). This algorithm is simple and interpretable. Although our multi-rider algorithm in \Cref{sec:multi-ride} does not use it directly, it can serve as a pruning step to improve the performance of any algorithm for the multi-rider FA problem: running it on each rider's notification set yields a near-optimal subset of that set, and the pruned sets remain disjoint.

% The FA setting is the main technical focus of the paper. We first study the single-rider problem and
% establish a Polynomial Time Approximation Scheme (PTAS). The algorithm relies on a structural
% characterization of optimal solutions, showing that only drivers within a narrow weight band around
% an endogenous threshold need to be selected carefully. We then extend our analysis to the general
% multi-rider setting. By relating the FA objective to a surrogate multinomial logit (MNL) valuation and
% combining a configuration LP with careful rounding and pruning, we obtain a constant-factor
% approximation algorithm.

\smallskip
\xhdr{Approximation algorithms for multiple riders (\Cref{sec:multi-ride}):}
We develop a deterministic polynomial-time algorithm that simultaneously achieves a $(1-1/e)$-approximation under FA and BA. Our approach uses a common ex-ante LP relaxation that upper-bounds the optimal welfare under both protocols. We first independently assign each driver to at most one rider according to the LP solution and attenuate notifications using the linear rule $a(y)=1-y/3$. Under FA, every rider--driver pair is matched with probability at least $(1-1/e)$ times its LP matching marginal. We then derandomize the algorithm using conditional expectations, preserving the welfare guarantee. Since BA yields at least the welfare of FA for any fixed notification sets, the same deterministic sets satisfy the approximation guarantee under both protocols. The common LP benchmark also establishes that the optimal FA welfare is at least $(1-1/e)$ times the optimal BA welfare. We show that this comparison is tight, even for a single rider when all drivers have the same acceptance probability.

\smallskip
\xhdr{Best accept: submodularity, demand oracle \& tractable special cases.} We exploit the structure of problem under BA to design a near-optimal polynomial-time \emph{demand oracle} for the particular monotone submodular function arising from BA. A demand oracle, given a vector of prices or costs for each driver, finds the subset that maximizes the value of the set (i.e., the expected matching score of the driver selected in this subset under BA) minus the sum of prices in that set. We show that this problem, after proper discretization, can be solved using a polynomial-time dynamic program. Using known results in the literature for SWM with access to the demand oracle~\citep{feige2010submodular}, we obtain an improved $(1 - 1/e+c)$-approximation for a small constant $c>0$. Finally, we prove that for the practically relevant special case where drivers share \emph{homogeneous acceptance probabilities}, the problem admits an exact polynomial-time solution via a linear programming formulation.

\smallskip
\xhdr{Simulations, interpretation \& practical takeaways.} We further run numerical simulations and verify the performance of our proposed approximation algorithms beyond worst-case instances. In particular, we measure their performance in synthetic, yet practical problem instances, as well as in real problem instances based on Lyft ride-sharing data. We observe consistent behavior compared to our theoretical results, and our algorithms perform even better in such instances. 

Our combination of theoretical and empirical analysis suggests three qualitative takeaways. First, there is a clear advantage in terms of single-cycle performance in NED vs. ED, under both FA and BA. Second, under FA, ``broadcasting more'' is not always better: adding a highly responsive but low-score driver can reduce expected welfare, and near-optimal sets use principled pruning (relying on certain threshold structure). Finally, under BA, broadcasting has diminishing returns and can be treated as a submodular welfare maximization problem, enabling the use of standard algorithmic tools (and stronger guaranties by leveraging efficient implementation of the demand oracle). These single-cycle insights complement Part~II~\ifblind\citep{LyftII2026companion_Blind}\else\citep{LyftII2026companion_NonBlind}\fi, which studies how such per-cycle choices interact with marketplace dynamics over time.

% Our paper connects to several strands of work in operations research and computer science. We briefly summarize the further related literature in Appendix~\ref{sec:related}. 
% \vspace{-1mm}

\subsection{Further Related Work}
\label{sec:related}
Our paper connects to several strands of work in operations research and computer science.

\smallskip
\xhdr{Assortment optimization under discrete choice models.}
Assortment optimization studies the selection of a subset of options to present to agents who then
choose according to a discrete-choice model. A large literature in operations research analyzes structural properties and approximation algorithms under parametric models such as MNL and its extensions, including settings
with capacity, inventory, visibility and robustness considerations
\citep{TalluriVanRyzin2004,RusmevichientongTopaloglu2012,Rusmevichientong2014RandomParams,
gallego2014constrained, desir2022capacitated,barre2025assortment,aouad2023stability}. Related work considers richer non-MNL
models, such as Markov-chain and ranking-based choice, which lead to different computational and
approximation phenomena~\citep{Feldman2014MarkovChoice,blanchet2016markov,asadpour2023sequential,niazadeh2021online,derakhshan2022product,agarwal2024misalignment,rieger2024quasi,aouad2023exponomial}, as well as consider-then-choose choice models~\citep{aouad2021assortment,asadpour2023sequential,niazadeh2021online}.
While FA valuation is indeed a special case of consider-then-choose, the positive algorithmic results for this class \emph{do not} apply, as the revenue function of the resulting consider-then-choose in this reduction could have exponentially many consideration sets.

Closer to us are two-sided assortment-optimization models for sequential matching markets, starting with the work of \citet{ashlagi2022assortmentplanning} and the \revcolor{subsequent work of \citet{aouad2023online,torrico2020multi}}; see also recent work on adaptivity gaps and approximation algorithms \citep{elhousni2024twosided} and \revcolor{on revenue maximization \citep{nissim2024revenue,ahmadnejadsaein2025adaptive}}. Similar to our BA welfare maximization problem, the match-maximization special case of these models---when one side nodes are assigned to the other side first by ignoring the other side's matching constraints, and then the other side nodes selects the final matches--- is also a special case of submodular welfare maximization.
Finally, our FA objective differs from classical MNL-based assortment purchase function or revenue function : it is non-monotone, non-submodular, and not even order-submodular (see, e.g., \cite{udwani2023submodular}).

\smallskip
\xhdr{Submodular welfare maximization.}
The BA protocol yields monotone submodular valuations, placing our multi-rider BA problem
within the submodular welfare maximization (SWM) framework. In the standard \emph{value-oracle}
model, the optimal approximation factor is $(1-1/e)$: it is achieved via the (randomized) continuous
greedy method \citep{vondrak2008optimal,CalinescuChekuriPalVondrak2011}, and it is essentially tight
under polynomial-time computation and/or polynomially many value queries \citep{KhotEtAl2008,MirrokniSchapiraVondrak2008}.
In stronger oracle models, SWM admits improved guarantees: with \emph{demand queries} one can
surpass the $(1-1/e)$ barrier---e.g. see  \cite{DobzinskiSchapira2006,feige2010submodular}---reflecting the
additional power of demand access (and the fact that hard instances for value-oracle lower bounds may encode demand queries of exponential complexity). We leverage the structure of BA valuations
to obtain improved approximation guarantees by designing efficient demand oracles. In general, our work draws ideas from the literature on monotone/non-monotone submodular maximization and extends this literature by considering FA valuations, which are neither monotone nor submodular~\citep{CalinescuChekuriPalVondrak2011,feige2011maximizing,buchbinder2015tight,niazadeh2020optimal}.

\smallskip
\xhdr{Ride-hailing, dynamic matching, and matching queues.}
A large literature studies dynamic matching in transportation and online platforms, including models
with waiting, time windows, and abandonment; see, e.g., \cite{ozkan2020dynamic,aouad2020dynamic,
ashlagi2019edge,amanihamedani2024improved,aveklouris2025matching} and references therein.  These papers typically
focus on dynamic policies and steady-state performance, often under binding allocations.
Our setting differs in that driver participation is explicitly stochastic (offers can be ignored/rejected),
and the platform may deliberately send overlapping invitations via \FEedit{NED} notifications. More importantly, the focus of our work is understanding the computational landscape of the single-cycle optimization problem, in contrast to the dynamic and long-term effects of notification mechanisms (which is studied in our companion paper \ifblind\cite{LyftII2026companion_Blind}\else\cite{LyftII2026companion_NonBlind} \fi.) 

\smallskip
\xhdr{BA/FA welfare maximization beyond ride-hailing.}
Broadcast-offer allocation with stochastic participation---where the platform chooses whom to solicit and allocation goes either to the first responder (FA) or the best among (timely) responders (BA)---also appears in food rescue and donation platforms \citep{BenadeAlptekinoglu2024RawlsianFoodRescue,LeeManshadiSaban2025WhoToOffer,ShiYuanLoLizarondoFang2020FoodRescue}, community first-responder dispatch \citep{HendersonVanDenBergJagtenbergLi2022VolunteersDispatch,DellaertSchlicherHillenaarJagtenberg2024CFRNetworks}, and spatial crowdsourcing systems that multicast tasks and then finalize among responders \citep{BasikGedikFerhatosmanogluWu2021CrowdsourcedDelivery}. Another possible application of our framework is in deceased-donor transplantation, where organ offers are sequentially (and sometimes
more broadly) extended to candidates/centers, whose acceptance behavior is heterogeneous and time-sensitive; see, for example
\cite{wey2017influence,husain2019association,agarwal2025choices}. Finally, a closely related BA instantiation arises in online advertising “header bidding,” where impressions are broadcast to multiple exchanges and the highest timely bid wins \citep{PachilakisPapadopoulosMarkatosKourtellis2019HeaderBidding,aqeel2020untangling}.

\revcolor{Our FA single-rider protocol is structurally analogous to the single-item last-minute offers model in \cite{cominetti2010optimal}. However, while their PTAS relies on an "instance-cloning" technique that alters customer probabilities, ours provides a simpler, more explicit, and faster algorithm. By exploiting an intrinsic threshold stability property that restricts near-optimal choices to a narrow middle-weight band, our approach uncovers vital structural insights into the optimal solution, specifically establishing sharp threshold rules for set selection. This yields an interpretable framework that serves as an embedded pruning criterion to eliminate harmful contention in the multi-rider setting.}

\revcolor{Closest to us are the concurrent work by \cite{liu2025recommend} and also \cite{mohammadimulti}. \cite{liu2025recommend} study a two-stage ``recommend-to-match'' problem for crowd-sourcing logistics/freight platforms under stochastic supplier rejections. Their formulation coincides with the Best-Accept objective in our model (up to an additional constraint that each request is recommended to at most some number of suppliers) and enforces the same supply-side exclusivity (each supplier receives at most one recommendation). They focus on tractable mathematical programming approaches, giving an exact MILP for the homogeneous-acceptance special case and proposing a mixed-integer exponential cone approximation with parametric performance bounds and extensive numerical evaluation. In contrast, we provide an \emph{exact polynomial-time} algorithm for the BA homogeneous-acceptance special case. Also, our work provides complexity results and worst-case approximation algorithms for both Best-Accept and First-Accept contention protocols in the general case, including a PTAS/constant-factor guarantees and improved approximation beyond $1-1/e$ for BA via demand-oracle methods.}

The model by \cite{mohammadimulti}, on the other hand, corresponds to our First-Accept objective but extends the setting to a more general multi-round problem where a task can be notified to workers over $T$ rounds. To approach this, they formulate a configuration LP and rely on the assumption that they have access to an oracle capable of solving it exactly. Given this optimal LP solution, they obtain a $\frac{1-e^{-3}}{3}$ approximation for the multi-round problem and a $\frac{1-e^{-2}}{2}$ approximation for the single-round problem. In contrast, while our work focuses on the single-round ($T=1$) setting, we do not assume the configuration LP can be solved exactly and instead provide unconditional, end-to-end polynomial-time algorithms. Specifically, we establish a deterministic $(1-1/e)$-approximation for the general multi-rider problem under both FA and BA, which is larger than the $\frac{1-e^{-2}}{2}$ factor above for the single-round problem, and a PTAS for the single-rider FA problem. Furthermore, for the homogeneous-driver setting where acceptance probabilities and rewards are ride-dependent ($p_{ij} = p_i, w_{ij}=w_i$), we provide an exact polynomial-time algorithm for the actual underlying integer optimization problem, whereas their results yield an exact algorithm only for the configuration LP relaxation---and therefore approximation algorithms for the original problem.

\subsection{Organization}
The remainder of this paper is organized as follows. In Section~\ref{sec:prelim}, we formally define the ride-hailing setting, the notification protocols, and the welfare maximization objective. Section~\ref{sec:single-ride} is dedicated to the single-rider problem, where we present a PTAS for the First Acceptance (FA) mechanism. Section~\ref{sec:multi-ride} extends the analysis to the general multi-rider setting under both BA and FA, where we derive a constant-factor approximation algorithm. In Section~\ref{sec:BA}, we then focus on the Best Acceptance (BA) mechanism, presenting an exact polynomial-time algorithm for ride-dependent acceptance probabilities, along with an improved approximation guarantee for the general multi-rider case. In Section~\ref{sec:numerical}, we compare our approximation algorithms with the optimal and greedy benchmarks, as well as the Exclusive Dispatch (ED) algorithm, using both synthetic instances and instances calibrated from real-world Lyft data. Finally, we conclude with a summary of our results and a discussion of open problems in Section~\ref{sec:conclusion}.

\section{Preliminaries}
\label{sec:prelim}
Ride-hailing platforms typically run dispatch in short decision cycles.  In each cycle, the platform observes a batch of unmatched rider requests and a batch of available drivers and must decide which drivers to \emph{notify} about which requests. Due to safety and operational considerations, a driver should not receive conflicting offers at the same time, so within a cycle each driver can be notified for at most one rider.  Our focus in this paper is the resulting single-cycle optimization problem, that is, how to send (possibly non-exclusive) notifications to optimize the quality of the eventual matching between riders and drivers. Below, we formalize this problem.

\paragraph{Riders, drivers, and match primitives.}
Let $\RiderSet=[\NumRiders]$ be the set of riders and $\DriverSet=[\NumDrivers]$ be the set of drivers.
For each rider--driver pair $(r,d)$, we are given:
\begin{itemize}
    \item a \emph{matching score} (also referred to as \emph{weight}) $\Weight{r}{d}\in[0,\wmax]$, capturing the platform's value from matching
    rider $r$ with driver $d$ in this cycle (e.g., a normalized function of ETA, pickup distance, and other
    features used by the dispatch model), and
    \item an \emph{acceptance probability} $\AccProb{r}{d}\in[0,1]$, the probability that driver $d$ accepts
    rider $r$'s request if notified (e.g., predicted by a driver-response model). 
\end{itemize}
We refer to $1-\AccProb{r}{d}$ as the probability of \emph{\FEedit{rejection}} throughout.\footnote{Estimated acceptance probabilities can vary across drivers for the same ride for different reasons. For example, a driver at a longer distance from a ride is more likely to reject.} We assume independent accept/reject decisions across drivers.  Also, we assume that drivers' response times to notifications are independent from accept/reject decisions, and, for simplicity of exposition, we assume that they are independent and identically distributed across all drivers and notifications.
\footnote{
% Our results extend in a straightforward fashion to the setting with heterogeneity in drivers' response times across drivers when response time distributions are known and memory-less (i.e., drawn from exponential distributions); however, 
We do not have any statistical evidence on 
% the degree of 
such heterogeneity in drivers' response times---and whether it is of first-order significance---in practice based on our proprietary Lyft data.}

\paragraph{Notification sets \& feasibility.}
A (non-exclusive) notification policy chooses disjoint subsets $(S_1,\dots,S_{\NumRiders})$ where
$S_r\subseteq \DriverSet$ is the set of drivers notified about rider $r$ and $S_r\cap S_{r'}=\emptyset$ for
$r\neq r'$. A matching algorithm is therefore defined by two components: (i) a notification policy that selects the subsets $S_r$ and (ii) a contention resolution protocol that determines which driver \emph{wins} when multiple notified drivers accept the same ride.

\paragraph{Platform's welfare maximization problem.} Fixing a contention resolution protocol, each rider $r$ induces a valuation function $F_r:2^{\DriverSet}\to \mathbb{R}_{\ge 0}$, where $\Fi{r}{S}$ is defined as the expected matching score of the eventual winning driver under this protocol, when $S$ is the set of drivers notified for rider $r$.  Given these valuation functions, the platform's \emph{single-cycle welfare} is defined as $\sum_{r\in\RiderSet}\Fi{r}{S_r}$. Our goal is to design a notification policy that maximizes this welfare subject to feasibility.

To study the above welfare maximization problem, we consider two contention resolution protocols that capture a basic speed--quality
trade-off in practice: \emph{Best Accept} and \emph{First Accept}.

\subsection{Best Accept Contention Resolution Protocol}
\label{sec:BA-def}
Under the Best Accept (BA) protocol, the platform seeks to maximize match quality. If a set of drivers $S$ is notified, the platform waits to receive all of their responses and then selects the highest-scoring driver among those who accept.

\begin{definition}[Best Accept Valuation]
\label{def:BA}
Let $\AccRV{r}{d}\sim \Bern(\AccProb{r}{d})$ be independent acceptance indicators.  If at least one driver in
$S$ accepts, rider $r$ is matched to the driver
$d^*=\arg\max\{\Weight{r}{d}: d\in S,\ \AccRV{r}{d}=1\}$; otherwise the rider remains unmatched.
The expected reward is
\begin{align}
\Fi{r}{S}
:=\E\!\left[\max_{d\in S:\ \AccRV{r}{d}=1}\Weight{r}{d}\right],
\label{eq:Fi-BA-def}
\end{align}
where the maximum over an empty set is $0$.
\end{definition}
\begin{proposition}[Monotonicity and submodularity of BA valuations]
\label{prop:BAsubmodular}
The BA valuation function $\Fi{r}{\cdot}$ defined in
\eqref{eq:Fi-BA-def} is monotone increasing and submodular.
\end{proposition}

\begin{proofof}{Proof}
Fix a rider $r$. Let $A\subseteq \DriverSet$ denote the (random) set of drivers who accept when notified, i.e.,
$A=\{d\in\DriverSet:\AccRV{r}{d}=1\}$. For any fixed realization $A$, define
\[
g_A(S):=\max_{d\in S\cap A}\Weight{r}{d},
\]
with the convention $g_A(\emptyset)=0$. Then $g_A$ is monotone and submodular: the marginal gain from
adding a driver $d$ to $S$ is either $0$ (if the current maximum in $S$  exceeds $\Weight{r}{d}$ or if $d\notin A$) or it is
$\Weight{r}{d}-g_A(S)$, and this marginal can only decrease as the set $S$ grows. Finally, \eqref{eq:Fi-BA-def} can be written
as $\Fi{r}{S}=\E[g_A(S)]$. Expectations preserve monotonicity and submodularity, so $\Fi{r}{\cdot}$ is monotone submodular.
\end{proofof}

One consequence of this result is that for the single rider unconstrained problem, studied in \Cref{sec:single-ride}, it is optimal to notify all drivers. The multi-rider welfare maximization problem remains nontrivial, however, because each driver can be notified about at most one rider.

\subsection{First Accept Contention Resolution Protocol}
\label{sec:FA-def}
Under the First Accept (FA) protocol, the platform assigns the ride to the first driver who responds. As drivers' response times are i.i.d. and independent of accept/reject decisions, this is equivalent to selecting a \emph{uniformly random} driver among the subset of notified drivers who accept. This protocol models systems that prioritize fast confirmation and therefore do not wait to compare all acceptors.

\begin{definition}[First Accept Valuation]
\label{def:FA}
Let $\AccRV{r}{d}\sim \Bern(\AccProb{r}{d})$ be independent acceptance indicators.  If at least one driver in
notified set $S\subseteq \DriverSet$ accepts, rider $r$ is matched to $d^*\in\{d\in S,\ \AccRV{r}{d}=1\}$ uniformly at random; otherwise the rider remains unmatched. The expected reward for rider $r$ is
\begin{align}
\Fi{r}{S}
:= \sum_{d \in S} \Weight{r}{d}\;
\E\!\left[
\Ind{\sum_{k \in S} \AccRV{i}{k} \ge 1}\;
\frac{\AccRV{r}{d}}{\sum_{k \in S} \AccRV{i}{k}}
\right].
\label{eq:Fi-FA-def}
\end{align}
\end{definition}
In stark contrast to BA, the FA valuation function $\Fi{r}{\cdot}$ defined in \eqref{eq:Fi-FA-def} is generally neither monotone nor submodular. Adding a low-score driver can reduce welfare by increasing contention against higher-score drivers. Moreover, the marginal value of a driver can increase as the notification set grows, since the driver may dilute the contention created by even lower-score drivers. This non-monotonicity also implies that an optimal notification set need not be a prefix of the drivers ordered by score: it can be preferable to skip an intermediate-score driver in favor of a more reliable lower-score driver. The following example demonstrates all three properties.

\begin{example}[FA: non-monotone, not submodular, and non-prefix optima]
\label{ex:FA-properties}
Consider a single rider and four drivers with $w_1\ge w_2\ge w_3\ge w_4$:\footnote{We use integer weights for ease of computation. Dividing all weights by $10$ rescales the example so that $w_{r,d}\in[0,1]$ without changing any of the comparisons.}
\[
(w_1,p_1)=(10,\tfrac12),\qquad (w_2,p_2)=(5,\tfrac14),\qquad (w_3,p_3)=(4,1),\qquad (w_4,p_4)=(1,1).
\]
Under FA, the rider is matched to a uniformly random acceptor.

\emph{(i) Non-monotonicity.} $F^{\mathrm{FA}}(\{1\})=10\cdot\tfrac12=5$, whereas
\[
F^{\mathrm{FA}}(\{1,4\})=\tfrac12\cdot\tfrac{10+1}{2}+\tfrac12\cdot 1=\tfrac{13}{4}<5 .
\]
Driver~4 always accepts, so on the event that driver~1 accepts, the reliable but poor driver~4 wins half of the time.
Notifying an additional driver can therefore reduce welfare.

\emph{(ii) No submodularity.} The marginal value of driver~2 can increase with the set it is added to:
\[
F^{\mathrm{FA}}(\{2,3\})-F^{\mathrm{FA}}(\{3\})
=\Big(\tfrac14\cdot\tfrac{5+4}{2}+\tfrac34\cdot 4\Big)-4=\tfrac18,
\]
\[
F^{\mathrm{FA}}(\{2,3,4\})-F^{\mathrm{FA}}(\{3,4\})
=\Big(\tfrac14\cdot\tfrac{5+4+1}{3}+\tfrac34\cdot\tfrac{4+1}{2}\Big)-\tfrac52=\tfrac{5}{24}>\tfrac18 .
\]
When driver~4 is present, driver~2 also dilutes driver~4's share of the contention, which makes driver~2 more valuable.

\emph{(iii) The optimal set need not be a prefix.} Among the prefixes of the score order,
\[ F^{\mathrm{FA}}(\{1\})=5,\quad F^{\mathrm{FA}}(\{1,2\})=\tfrac{85}{16},\quad F^{\mathrm{FA}}(\{1,2,3\})=\tfrac{263}{48},\quad F^{\mathrm{FA}}(\{1,2,3,4\})=\tfrac{185}{48}, \]
whereas
\[ F^{\mathrm{FA}}(\{1,3\}) =\frac12\cdot\frac{10+4}{2}+\frac12\cdot 4 =\frac{11}{2}.\]
A direct evaluation of the remaining subsets shows that the best competing set is $\{1,2,3\}$, whose value is $\frac{263}{48}<\frac{11}{2}$. Thus, $\{1,3\}$ is the unique optimum over all $2^4$ subsets. In particular, the optimal set skips driver~$2$: driver~$3$ provides a reliable fallback when driver~$1$ declines, whereas the limited benefit from notifying driver~$2$ is outweighed by the contention it creates with driver~$1$.

\emph{(iv) Contrast with BA.} Under BA, the full prefix
$\{1,2,3,4\}$ is optimal and has value $\frac{57}{8}$. Driver~$4$ is
redundant, however, because driver~$3$ always accepts; consequently,
$\{1,2,3\}$ is also optimal.
\end{example}

Consequently, even the single-rider problem is nontrivial. We study this problem in \Cref{sec:single-ride}.

\subsection{$k$-Accept Contention Resolution Protocol}
We also consider a generalized contention resolution protocol, called $k$-Accept, that interpolates between FA and BA. Under this protocol, the platform notifies the drivers in $S$ and finalizes a match upon receiving the $k$-th acceptance, selecting the highest-weight driver among the first $k$ acceptors. If fewer than $k$ drivers accept, the platform selects the highest-weight acceptor once all responses have arrived or the response window has closed; if no driver accepts, the rider remains unmatched. The parameter $k$ controls the trade-off between matching speed and quality: smaller values allow earlier confirmation, while larger values permit comparison among more accepting drivers. Setting $k=1$ recovers FA and the valuation in \eqref{eq:Fi-FA-def}. Setting $k=\infty$, or any $k\geq|\mathcal{D}|$, ensures that all accepting drivers are considered and recovers BA and the valuation in \eqref{eq:Fi-BA-def}.

\subsection{NP-Hardness}
\label{sec:hardness}
% We end this section by showing that welfare maximization under either any $k$-accept protocols, including FA and BA, is \emph{strongly NP-hard}. First, we observe that for instances with homogeneous weights, the valuation functions are identical for any $k-$accept. Thus, the hardness result with homogeneous weights applies to all settings simultaneously.

% Concretely, consider the decision version of the problem in which the input also includes a rational threshold $W$, and one asks whether there exists a feasible allocation $(S_1,\dots,S_{\NumRiders})$ (disjoint notification sets) with $\sum_{r\in\RiderSet}\Fi{r}{S_r}\ \ge\ W.$
% We formally show the following theorem, with proof deferred to Appendix~\ref{appendix:hardness}, which justifies the search for approximation algorithms.

% \begin{theorem}[Strong NP-hardness of FA/BA welfare maximization]
% \label{thm:hardness}
% For each $k$, the welfare maximization under $k-$accept valuations is strongly NP-hard even under the following restrictions:
% \begin{itemize}
%   \item all riders are identical (i.e., $\AccProb{r}{d}$ and $\Weight{r}{d}$
%         do not depend on $r$),
%   \item all weights are unit: $\Weight{r}{d}\equiv 1$,
%   \item all probabilities are dyadic rationals of the form
%         $\AccProb{r}{d}=1-2^{-a_d}$ for integers $a_d\ge 1$.
% \end{itemize}
% Consequently, the general welfare maximization problem for the $k-$accept valuation class is strongly NP-hard and admits no FPTAS unless $\mathrm{P}=\mathrm{NP}$
% \end{theorem}

We end this section by showing that computing an optimal allocation is NP-hard under either FA or BA. When all weights equal one, the two protocols yield the same reward: one if at least one notified driver accepts, and zero otherwise. A hardness result for this restricted class therefore applies to both protocols.

\begin{theorem}[NP-hardness of FA/BA welfare maximization] \label{thm:hardness} Computing an optimal allocation under either FA or BA (or $k-$accept) is NP-hard, even with two identical riders ($\AccProb{r}{d} = p_d$) and unit weights, i.e. $\Weight{r}{d}\equiv 1$. \end{theorem}

The proof, deferred to Appendix~\ref{appendix:hardness}, proceeds by a reduction from \textsc{Product Partition}. This hardness result motivates the approximation algorithms developed in \Cref{sec:multi-ride}. We further show in \Cref{sec:single-ride} that the FA problem remains NP-hard even with a single rider, for which we develop the PTAS in Algorithm~\ref{alg:single}.

\section{Single-Rider NED Notifications}
\label{sec:single-ride}
In this section, we study the unconstrained \emph{single-rider} version of the welfare maximization problem. Under \emph{BA}, \Cref{prop:BAsubmodular} implies that the valuation is monotone increasing, so it is optimal to notify all drivers. The nontrivial single-rider problem therefore arises under \emph{FA}, where we seek to maximize the FA valuation function in \eqref{eq:Fi-FA-def} over all subsets of drivers without any constraints. Throughout this section, we fix rider~$r$ and drop $r$ from our notation by letting
$\wSingle{d}:=\Weight{r}{d}$, $\pSingle{d}:=\AccProb{r}{d}$, $X_d\sim\Bern(\pSingle{d})$, and $F(S):=\Fi{r}{S}$ for all $S\subseteq\DriverSet$.

The main structural difficulty is that adding a low-weight driver can reduce the probability that a high-weight driver is selected, since the winner under FA is uniform among the accepting drivers. Consequently, $F(\cdot)$ is generally neither monotone nor submodular, as demonstrated in \Cref{ex:FA-properties}. The same example shows that, despite the similarity between the FA valuation and the standard Multinomial Logit (MNL) revenue function~\citep{TalluriVanRyzin2004}, an optimal FA notification set need not be ordered by weight: it can skip an intermediate-weight driver while including a lower-weight one. This contrasts with classical unconstrained assortment optimization under MNL, for which a revenue-ordered optimal assortment always exists and can be found by a simple linear search.

This lack of structure reflects a more fundamental computational difficulty: finding an optimal notification set under FA is NP-hard.

\begin{theorem}[NP-hardness of single-rider FA welfare maximization]
\label{thm:NPhard}
The single-rider welfare maximization problem under FA is NP-hard.
\end{theorem}

Despite the above challenges, we present Algorithm~\ref{alg:single} as our main result in this section, which provides a Polynomial-Time Approximation Scheme (PTAS) for $\max_{S\subseteq\DriverSet}\Fi{}{S}$. 

\medskip
\begin{algorithm}[htb]
\caption{Single-ride FA approximation (Algorithm 1)}
\label{alg:single}
\KwIn{Weights $\{\wSingle{d}\}_{d\in\DriverSet}$, probabilities $\{\pSingle{d}\}_{d\in\DriverSet}$, accuracy parameter $\del>0$}
\KwOut{A set $\hat S\subseteq\DriverSet$}

\smallskip
Sort so that $\wSingle{1}\ge \wSingle{2}\ge \cdots \ge \wSingle{\NumDrivers}$\;
\For{$k\gets 1$ \KwTo $\NumDrivers-1$}{

\smallskip
  \tcp{Guess: $\wSingle{k+1}\le \tauTilde \le \wSingle{k}$.}
  $\HighSet \gets \{d\in\DriverSet:\ \wSingle{d}\ge \wSingle{k}\}~~,~~\LowSet \gets \{d\in\DriverSet:\ \wSingle{d}<\wSingle{k+1}/3\}~~,~~\MidSet \gets \{d\in\DriverSet:\ \wSingle{k+1}/3 \le \wSingle{d}\le \wSingle{k+1}\}$\;
  $\HatL \gets \left\lceil \frac{\ln 3}{\ln(1+\del)}\right\rceil$\;
  \For{$\ell\gets 1$ \KwTo $\HatL$}{
    $\MidSetBucket{\ell} \gets \left\{d\in\MidSet:\ 
      \wSingle{d}\in\Bigl[\tfrac{\wSingle{k+1}}{3}(1+\del)^{\ell-1},\ \tfrac{\wSingle{k+1}}{3}(1+\del)^{\ell}\Bigr)\right\}$\;
  }
  
  \smallskip
  \tcp{Enumerate over count vectors.}
  \ForEach{$\vec n=(n_1,\dots,n_{\HatL})\in\{0,1,\dots,\NumDrivers\}^{\HatL}$}{
    In each bucket $\MidSetBucket{\ell}$, pick the $n_\ell$ drivers with largest $\pSingle{d}$; let $\SMid(\vec n)$ be their union\;
    Evaluate $\Fi{}{\HighSet \cup \SMid(\vec n)}$\;
  }
  Let $\vec n^*\in\{0,1,\dots,\NumDrivers\}^{\HatL}$ maximize $F({\HighSet \cup \SMid(\vec n)})$ and set $S^{(k)}\gets \HighSet\cup \SMid(\vec n^*)$\;
}
\KwRet{$\hat S \gets \argmax_{k\in\{1,\dots,\NumDrivers-1\}} \Fi{}{S^{(k)}}$}\;
\end{algorithm}
\medskip

At a high level, our algorithm exploits a threshold structure that is specific to the FA objective and matches the operational intuition of \emph{include all very good drivers, exclude all very bad drivers, and carefully tune
the middle}. Specifically, it ``guesses'' the (unknown) quality threshold of an optimal solution, includes all drivers above that threshold, discards all drivers far below it, and discretizes the remaining middle band into buckets.  Within each bucket, the algorithm keeps the drivers with the largest acceptance probabilities. Formally, we show the following theorem.

\begin{theorem}[PTAS for the {single-rider FA problem}]
\label{thm:alg1}
Given oracle access to $\Fi{}{\cdot}$, where $\Fi{}{\cdot}$ belongs to the class of FA valuation functions defined in Definition~\ref{def:FA}, Algorithm~\ref{alg:single} with any accuracy parameter $\delta>0$ runs in time $\NumDrivers^{O(1/\del)}$ and returns $\hat S$ such that
\[
\Fi{}{\hat S}\ \ge\ (1-\del)\cdot \max_{S\subseteq\DriverSet}\Fi{}{S}.
\]
\end{theorem}

We analyze Algorithm~\ref{alg:single} through a sequence of structural lemmas. The first lemma gives an integral representation that makes the objective amenable to algebraic manipulation.

\begin{lemma}[Integral representation and multilinearity]
\label{lem:int-rep}
For any $S\subseteq\DriverSet$,
\begin{equation}
F(S)
=
\sum_{d\in S}\wSingle{d}\pSingle{d}
\int_0^1\ \prod_{k\in S\setminus\{d\}}\bigl(1-\pSingle{k}+\pSingle{k}t\bigr)\,dt.
\label{eq:int-rep}
\end{equation}
In particular, $F$ is linear in the weights $\{\wSingle{d}\}$ and multilinear in the probabilities
$\{\pSingle{d}\}$.
\end{lemma}

\begin{proofof}{Proof}
Starting from~\eqref{eq:Fi-FA-def} and using the single-rider shorthand,
\[
F(S)=\sum_{d\in S}\wSingle{d}\E\left[\Ind{\sum_{k\in S}X_k\ge 1}\cdot \frac{X_d}{\sum_{k\in S}X_k}\right]
=\sum_{d\in S}\wSingle{d}\pSingle{d}\E\left[\frac{1}{1+\sum_{k\in S\setminus\{d\}}X_k}\right].
\]
Apply the identity $\frac{1}{1+z}=\int_0^1 t^z\,dt$ for $z\ge 0$ and then independence:
\begin{align}\label{eq:FAidentity}
\E\left[\frac{1}{1+\sum_{k\in S\setminus\{d\}}X_k}\right]
=\int_0^1\E\left[t^{\sum_{k\in S\setminus\{d\}}X_k}\right]dt
=\int_0^1\ \prod_{k\in S\setminus\{d\}}\E[t^{X_k}]\,dt
=\int_0^1\ \prod_{k\in S\setminus\{d\}}\bigl(1-\pSingle{k}+\pSingle{k}t\bigr)\,dt.
\end{align}
Substituting back gives~\eqref{eq:int-rep} and finishes the proof.
\end{proofof}

The second lemma states that even though $F(\cdot)$ is non-monotone in general, it is always safe to include the best driver (recall Example~\ref{ex:FA-properties}). 
\begin{lemma}[The maximum-weight driver is always included]
\label{lem:max-weight}
Let $d\notin S$ be a driver with $\wSingle{d}\ge \max_{d\in S}\wSingle{d}$. Then $F(S\cup\{d\})\ge F(S)$.
In particular, there exists an optimal set $S^*\in\argmax_S F(S)$ that contains a driver of maximum weight.
\end{lemma}

\begin{proofof}{Proof}
Condition on whether driver $d$ accepts:
\[
F(S\cup\{d\})
=(1-\pSingle{d})F(S) + \pSingle{d}\cdot F\bigl(S\cup\{d\}\mid X_d=1\bigr).
\]
When $X_d=1$, the expected weight obtained by selecting uniformly among $d$ and the  acceptors in $S$ is at least the expected weight obtained by selecting uniformly only among the acceptors in $S$, since $\wSingle{d}$ is the maximum weight in $S\cup\{d\}$. Thus $F(S\cup\{d\}\mid X_d=1)\ge F(S)$ and the claim follows.
\end{proofof}
A central object in our analysis is the (set-dependent) weight threshold $\TauOf{S}$, which is defined in the following lemma.  Informally,
$\TauOf{S}$ is the minimum weight a new driver must have so that adding them to $S$ is beneficial under FA, or in other words, $\TauOf{S}$ is the break-even score that compensates for the extra contention created by adding one more potential acceptor to the pool

\begin{lemma}[Threshold for positive marginal value]
\label{lem:threshold}
Fix a subset $S\subseteq\DriverSet$ of drivers and consider adding a driver $d\notin S$ with parameters
$(\wSingle{d},\pSingle{d})$. There exists a threshold $\TauOf{S}$ depending only on $S$ (i.e., on $\{(\wSingle{d},\pSingle{d})\}_{d\in S}$)
such that
\[
F(S\cup\{d\})-F(S)\ge 0\ \Longleftrightarrow\ \wSingle{d}\ge \TauOf{S}.
\]
Moreover, the marginal value $F(S\cup\{e\})-F(S)$ is
linear in $\pSingle{d}$ with zero intercept when fixing other variables and the threshold $\TauOf{S}$ has the following closed form:
\begin{equation}
    \label{eq:threshold}
    \TauOf{S}:=\frac{\int_0^1(1-t)\sum_{d\in S}\wSingle{d}\pSingle{d}\prod_{k\in S\setminus\{d\}}\bigl(1-\pSingle{k}+\pSingle{k}t\bigr)\,dt}{\int_0^1\prod_{k\in S}\bigl(1-\pSingle{k}+\pSingle{k}t\bigr)\,dt}~.
\end{equation}

\end{lemma}

\begin{proofof}{Proof}
Start from the integral representation in Lemma~\ref{lem:int-rep}.
For any set $S$ and any $d\notin S$ we have:
\revcolor{\begin{align*}
F(S\cup\{d\})
&=\sum_{j\in S}\left(\wSingle{j}\pSingle{j}
\int_0^1\bigl(1-\pSingle{d}+\pSingle{d}t\bigr)
\prod_{k\in S\setminus\{j\}}\bigl(1-\pSingle{k}+\pSingle{k}t\bigr)\,dt\right)
+\wSingle{d}\pSingle{d}\int_0^1\prod_{k\in S}\bigl(1-\pSingle{k}+\pSingle{k}t\bigr)\,dt.
\end{align*}}
Subtracting $F(S)$ and simplifying yields
\[
F(S\cup\{d\})-F(S)=\pSingle{d}\cdot\Bigl(\wSingle{d}\,B(S)-A(S)\Bigr),
\]
where we have:
\[
B(S):=\int_0^1\prod_{k\in S}\bigl(1-\pSingle{k}+\pSingle{k}t\bigr)\,dt,
\qquad
A(S):=\int_0^1(1-t)\sum_{d\in S}\wSingle{d}\pSingle{d}\prod_{k\in S\setminus\{d\}}\bigl(1-\pSingle{k}+\pSingle{k}t\bigr)\,dt.
\]
Note that $B(S)>0$ always. If some $\pSingle{k}<1$ then the integrand is strictly positive for all $t\in[0,1]$,
and if all $\pSingle{k}=1$ then $B(S)=\int_0^1 t^{|S|}dt=1/(|S|+1)>0$.
Thus the sign of the marginal depends only on whether $\wSingle{d}\ge A(S)/B(S)$.
Therefore we define $\TauOf{S}:=A(S)/B(S)$.
Finally, the expression above shows the marginal is linear in $\pSingle{d}$ with zero intercept when fixing other variables. 
\end{proofof}

 % The fourth and last structural lemma (full proof in Appendix~\ref{appendix:bound-threshold}) 
The fourth and last structural lemma is the technical step behind the ``$\tau/3$'' cutoff in Algorithm~\ref{alg:single}, and the key technical lemma that makes the single-ride problem under FA amenable to a polynomial-time approximation scheme (PTAS): if a driver has nonnegative marginal against a set $T$, then after adding it the weight threshold defined in Lemma~\ref{lem:threshold} cannot jump by more than a constant factor.

\begin{lemma}[Threshold stability]
\label{lem:bound-threshold}
Let $T\subseteq\DriverSet$ be a subset of drivers and suppose that driver $d\notin T$.  If $\wSingle{d}\ge \TauOf{T}$, then have that $\TauOf{T\cup\{d\}}\le 3\wSingle{d}.$
\end{lemma}

% \begin{proof}[Proof sketch]
% Write the integrals defining $\TauOf{\cdot}$ in terms of the auxiliary function
% $g_T(t):=\prod_{k\in T}(1-\pSingle{k}+\pSingle{k}t)$, which is increasing on $t\in[0,1]$.
% Using that $g_T$ is increasing, one can show
% $\int_0^1 (1-t)g_T(t)\,dt \le \tfrac{1}{2}\int_0^1 g_T(t)\,dt$ and
% $\int_0^1 t\,g_T(t)\,dt \ge \tfrac{1}{2}\int_0^1 g_T(t)\,dt$.
% Combining these inequalities with the assumption $\wSingle{d}\ge \TauOf{T}$ yields
% $\TauOf{T\cup\{d\}}\le 3\wSingle{d}$. \RN{Does the new proof look good?}
% \end{proof}
\begin{proofof}{Proof}
Fix $T\subseteq\DriverSet$ and $d\notin T$.
Define
\[
g_T(t):=\prod_{k\in T}\bigl(1-\pSingle{k}+\pSingle{k}t\bigr),
\qquad
\revcolor{h_T(t):=\sum_{j \in T}\wSingle{j}\pSingle{j}\prod_{k\in T\setminus\{j\}}\bigl(1-\pSingle{k}+\pSingle{k}t\bigr)}.
\]
Then $B(T)=\int_0^1 g_T(t)\,dt$ and $A(T)=\int_0^1 (1-t)h_T(t)\,dt$. Now, let $S:=T\cup\{d\}$.  Since $(1-p_d+p_d t)\le 1$ for $t\in[0,1]$, we have
\begin{align*}
A(S)
&=\int_0^1(1-t)\Bigl[(1-p_d+p_d t)\,h_T(t) + w_d p_d\,g_T(t)\Bigr]\,dt\\
&\le \int_0^1(1-t)h_T(t)\,dt\ + w_d p_d\int_0^1(1-t)g_T(t)\,dt\\
&=A(T)\ + w_d p_d\int_0^1(1-t)g_T(t)\,dt.
\end{align*}
Similarly,
\[
B(S)=\int_0^1 (1-p_d+p_d t)g_T(t)\,dt=(1-p_d)B(T)+p_d\int_0^1 t\,g_T(t)\,dt.
\]

Each factor $(1-\pSingle{k}+\pSingle{k}t)$ is increasing in $t$, hence $g_T(t)$ is increasing on $[0,1]$.
For any nonnegative increasing function $g$ on $[0,1]$, Chebyshev's integral inequality implies
\[
\int_0^1 t\,g(t)\,dt\ \ge\ \left(\int_0^1 t\,dt\right)\left(\int_0^1 g(t)\,dt\right)=\frac{1}{2}\int_0^1 g(t)\,dt.
\]
Applying this to $g=g_T$ gives
\begin{equation}
\int_0^1 t\,g_T(t)\,dt\ \ge\ \frac{1}{2}B(T)
\qquad\text{and}\qquad
\int_0^1 (1-t)\,g_T(t)\,dt
=B(T)-\int_0^1 t\,g_T(t)\,dt\ \le\ \frac{1}{2}B(T).
\label{eq:gt-half}
\end{equation}
Using the inequality in ~\eqref{eq:gt-half} and $p_d\le 1$,
\[
A(S) \le A(T)+ w_dp_d\cdot \frac{1}{2}B(T) \le A(T)+ w_d\cdot \frac{1}{2}B(T).
\]
By assumption, $w_d\ge \TauOf{T}=A(T)/B(T)$, so $A(T)\le w_d B(T)$.  Therefore
\[
A(S) \le w_d B(T)+\frac{1}{2}w_d B(T)=\frac{3}{2}w_d B(T).
\]
On the other hand, by~\eqref{eq:gt-half},
\[
B(S)=(1-p_d)B(T)+p_d\int_0^1 t\,g_T(t)\,dt\ \ge\ (1-p_d)B(T)+p_d\cdot \frac{1}{2}B(T)\ \ge\ \frac{1}{2}B(T).
\]
Combining the two bounds yields
\[
\TauOf{S}=\frac{A(S)}{B(S)}\ \le\ \frac{\frac{3}{2}w_d B(T)}{\frac{1}{2}B(T)}=3w_d,
\]
which finishes the proof of our desired inequality.
\end{proofof}

% \begin{proofof}{Proof sketch}
% Recall the definition $\tau(S) = A(S)/B(S)$. We observe that for any added element $d$, the denominator satisfies the lower bound $B(T\cup\{d\}) \geq \frac{1}{2}B(T)$. Furthermore, invoking the assumption that $w_d \ge \tau(T)$, the numerator is bounded by $A(T\cup\{d\}) \leq \frac{3}{2}w_d B(T)$. Substituting these inequalities into the ratio yields:
% \begin{align*}
%     \tau(T\cup\{d\}) = \frac{A(T\cup\{d\})}{B(T\cup\{d\})} \leq \frac{\frac{3}{2}w_dB(T)}{\frac{1}{2}B(T)} = 3w_d.
% \end{align*}
% \end{proofof}

Let $S^{*}\in\argmax_S F(S)$ be an optimal set and define $\Tau^{*}:=\TauOf{S^{*}}$. Putting our lemmas together, we can obtain a structural characterization of the optimal set. In particular, drivers fall into three groups around $\Tau^{*}$: all drivers above $\Tau^{*}$ must be
included, drivers far below $\Tau^{*}$ are excluded, and only a middle band requires careful choice. Formally, we show the following proposition.

\begin{proposition}[Three-way structure of an optimal solution]
\label{lem:threeway}
Let $S^{*}\in\argmax_S F(S)$ be an optimal set and define $\Tau^{*}:=\TauOf{S^{*}}$.
Then:
\begin{enumerate}[label=(\roman*)]
  \item If $d\notin S^{*}$ and $\wSingle{d}>\Tau^{*}$, then $F(S^{*}\cup\{d\})>F(S^{*})$.
        In particular, every driver with weight strictly larger than $\Tau^{*}$ must belong to $S^{*}$.
  \item Every driver in $S^{*}$ has weight at least $\Tau^{*}/3$.
\end{enumerate}
\end{proposition}

\begin{proofof}{Proof}
Part~(i) follows immediately from Lemma~\ref{lem:threshold}: if $\wSingle{d}>\TauOf{S^{*}}$ then
the marginal of adding $d$ is strictly positive, contradicting optimality. To see the proof of Part~(ii), fix $j\in S^{*}$ and let $S:=S^{*}\setminus\{j\}$. By optimality, removing $j$ cannot improve the value,
so $F(S^{*})-F(S)\ge 0$. Therefore, by Lemma~\ref{lem:threshold}, $\wSingle{j}\ge\TauOf{S}$. Invoking Lemma~\ref{lem:bound-threshold} yields $\wSingle{j}\ge \TauOf{S^{*}}/3=\Tau^{*}/3$.
\end{proofof}

Using this structural characterization, we are now ready to prove the main result of this section.
\begin{proofof}{Proof of \Cref{thm:alg1}}
We assume without loss of generality that $\pSingle{d}>0$ for every driver $d$, since drivers with $\pSingle{d}=0$ never accept and do not affect $F$.
Let $S^{*}\in\argmax_S F(S)$ be an optimal set of maximum cardinality, and let $\Tau^{*}:=\TauOf{S^{*}}$.
Sort weights so that $\wSingle{1}\ge\cdots\ge \wSingle{\NumDrivers}$.
By the maximality of $S^{*}$, we have $F(S^{*}\cup\{d\})<F(S^{*})$ for every $d\notin S^{*}$; hence, by \Cref{lem:threshold}, every driver with weight at least $\Tau^{*}$ belongs to $S^{*}$. Moreover, $S^{*}\neq\emptyset$ (since $F(\{d\})=\wSingle{d}\pSingle{d}\ge 0=F(\emptyset)$ for every $d$), and applying \Cref{lem:max-weight,lem:threshold} to a hypothetical driver $d\notin S^{*}$ with $\wSingle{d}=\max_{j\in S^{*}}\wSingle{j}$ yields $\Tau^{*}\le\max_{j\in S^{*}}\wSingle{j}\le\wSingle{1}$. If $\Tau^{*}\le\wSingle{\NumDrivers}$, then every driver belongs to $S^{*}$, i.e., $S^{*}=\DriverSet$; in this case, the iteration $k=\NumDrivers-1$ of Algorithm~\ref{alg:single} has $\HighSet\cup\MidSet=\DriverSet$, and choosing $n_\ell=|\MidSetBucket{\ell}|$ for all $\ell$ gives $\SMid(\vec n)=\MidSet$, so $F(\hat S)\ge F(\DriverSet)=F(S^{*})$. Otherwise, $\wSingle{\NumDrivers}<\Tau^{*}\le\wSingle{1}$, and we can
choose $k\in\{1,\dots,\NumDrivers-1\}$ such that
\begin{equation}
\wSingle{k}\ge \Tau^{*} > \wSingle{k+1}.
\label{eq:guess-k}
\end{equation}
Consider the iteration of Algorithm~\ref{alg:single} corresponding to this $k$. We prove our theorem step-by-step:

\paragraph{Step 1: drivers above and below the threshold.}
As shown above, every driver with weight at least $\Tau^{*}$ belongs to $S^{*}$. Since weights are sorted and~\eqref{eq:guess-k} holds, this implies $\HighSet\subseteq S^{*}$.
Similarly, by Part~(ii) of \revcolor{\Cref{lem:threeway}}, every driver in $S^{*}$ has weight at least $\Tau^{*}/3$.
Since $\Tau^{*}>\wSingle{k+1}$, we have $\Tau^{*}/3>\wSingle{k+1}/3$, so no element in
$\LowSet=\{j:\wSingle{j}<\wSingle{k+1}/3\}$ can belong to $S^{*}$.
Hence the optimal set has the form
\[
S^{*} = \HighSet\ \cup\ \bigl(S^{*}\cap \MidSet\bigr),
\]
and the algorithm only needs to approximate the choice inside $\MidSet$.

\paragraph{Step 2: bucketing and rounding.}
Partition $\MidSet$ into $\HatL=\lceil\ln 3/\ln(1+\del)\rceil$ buckets as in the algorithm.
For each bucket $\MidSetBucket{\ell}$, all weights lie in an interval of multiplicative width $(1+\del)$.
Define rounded-down weights $\tilde w_j$ by setting $\tilde w_j$ to the lower endpoint of its bucket.
Then for every $j\in\MidSet$ we have
$\tilde w_j\le \wSingle{j}\le (1+\del)\tilde w_j$.
By Lemma~\ref{lem:int-rep}, $F(\cdot)$ is linear in the weights, hence for any fixed set $U\subseteq\MidSet$,
\begin{equation}
\tilde F(\HighSet\cup U)\ \le\ F(\HighSet\cup U)\ \le\ (1+\del)\,\tilde F(\HighSet\cup U),
\label{eq:rounding}
\end{equation}
where $\tilde F$ denotes the valuation computed with rounded-down weights.

\paragraph{Step 3: guessing bucket counts and choosing top probabilities.}
Let $U^{\sharp}\in\argmax_{U\subseteq\MidSet}\tilde F(\HighSet\cup U)$, and let $n^{\sharp}_\ell:=|U^{\sharp}\cap\MidSetBucket{\ell}|$ for every $\ell$.
Algorithm~\ref{alg:single} enumerates all count vectors in $\{0,1,\dots,\NumDrivers\}^{\HatL}$, so it considers $\vec n^{\sharp}$.
Fix a bucket $\MidSetBucket{\ell}$; under the rounded weights, all drivers in this bucket have the same weight $\tilde w$.
Suppose that $a\in U^{\sharp}\cap\MidSetBucket{\ell}$ and $b\in\MidSetBucket{\ell}\setminus U^{\sharp}$ satisfy $\pSingle{b}>\pSingle{a}$, and let $T:=\HighSet\cup(U^{\sharp}\setminus\{a\})$.
By \Cref{lem:threshold} (see its proof) applied to the rounded weights, $\tilde F(T\cup\{c\})-\tilde F(T)=\pSingle{c}\,\bigl(\tilde w\,B(T)-\tilde A(T)\bigr)$ for $c\in\{a,b\}$, where $B(T)>0$ and $\tilde A(T)$ are the denominator and the numerator in \eqref{eq:threshold}, the latter computed with the rounded weights.
By the optimality of $U^{\sharp}$, removing $a$ cannot increase $\tilde F$, so $\tilde w\,B(T)-\tilde A(T)\ge 0$ and hence $\tilde F(T\cup\{b\})\ge\tilde F(T\cup\{a\})$; that is, replacing $a$ by $b$ preserves optimality.
Since each such swap increases the total probability of the selected drivers in the bucket, repeating such swaps in every bucket yields, after finitely many steps, an optimal set whose intersection with each bucket $\MidSetBucket{\ell}$ consists of $n^{\sharp}_\ell$ drivers with the largest probabilities in that bucket. This set has the same value of $\tilde F$ as the set $\SMid(\vec n^{\sharp})$ constructed by the algorithm, since drivers in the same bucket with equal probabilities are interchangeable under the rounded weights. Consequently,
\[\tilde F\bigl(\HighSet\cup\SMid(\vec n^{\sharp})\bigr)\ =\ \max_{U\subseteq\MidSet}\tilde F(\HighSet\cup U)\ \ge\ \tilde F\bigl(\HighSet\cup (S^{*}\cap \MidSet)\bigr)=\tilde F(S^{*}).\]

\paragraph{Step 4: putting everything together.} Let $S^\dagger:=\HighSet\cup\SMid(\vec n^{\sharp})$ denote the candidate set corresponding to the count vector $\vec n^{\sharp}$ from Step~3.
By Step~3, $\tilde F(S^\dagger)\ge \tilde F(S^{*})$.
Since the true weights dominate the rounded ones, we also have $F(S^\dagger)\ge \tilde F(S^\dagger)$.
Finally, Algorithm~\ref{alg:single} outputs $\hat S$ maximizing $F(\cdot)$ over all its candidates, so
$F(\hat S)\ge F(S^\dagger)$.
Combining these with~\eqref{eq:rounding} applied to $U=S^{*}\cap\MidSet$ yields
\[
F(\hat S)\ \ge\ F(S^\dagger)\ \ge\ \tilde F(S^\dagger)\ \ge\ \tilde F(S^{*})\ \ge\ \frac{1}{1+\del}F(S^{*})\ \ge\ (1-\del)F(S^{*}),
\]
which is the claimed approximation.

\paragraph{Running time analysis.} The running time is dominated by enumerating $\vec n\in\{0,1,\dots,\NumDrivers\}^{\HatL}$ with
$\HatL=\Theta(1/\del)$, giving $\NumDrivers^{O(1/\del)}$ evaluations.
\end{proofof}

\begin{remark}
    While \eqref{eq:Fi-FA-def} is a summation over exponentially many terms and \Cref{eq:int-rep} is an integral representation, the value of $F(\cdot)$ can be calculated efficiently. Under FA, driver $d\in S$ wins with probability $1/(j+1)$ whenever exactly $j$ other drivers
in $S$ accept, regardless of their scores. Hence
    \begin{align*}
        F(S) = \sum_{d\in S} w_dp_d \sum_{j=0}^{|S|-1} \frac{\mathbb{P}[j \text{other drivers accept}]}{j+1},
    \end{align*}
    where the probabilities can be computed using a simple DP.
\end{remark}

% \begin{proof}
% Apply Algorithm~\ref{alg:single} to the instance obtained by restricting attention to the ground set $S$.
% Proposition~\ref{prop:alg1} (applied to the restricted instance) guarantees a set $\hat S\subseteq S$ with
% \[
% \Fi{r}{\hat S}\ \ge\ (1-\del)\cdot \max_{U\subseteq S}\Fi{r}{U}=(1-\del)\,\Fibar{i}{S}.
% \]
% \end{proof}

\section{Multi-Rider NED Notifications}
\label{sec:multi-ride}
In this section, we study the single-cycle welfare maximization problem with multiple riders. The platform chooses a notification set $S_r\subseteq\mathcal{D}$ for each rider $r\in\mathcal{R}$, subject to the restriction that each driver receives at most one notification. The objective is to maximize the total expected matching score:
\begin{align}
\label{eq:mainP}\tag{$\mathcal{P}$}
\max_{(S_r)_{r\in\mathcal{R}}}\quad
&\sum_{r\in\mathcal{R}}F_r(S_r)
&
\text{s.t.}\quad
&S_r\subseteq\mathcal{D}
&&\forall r\in\mathcal{R}, \nonumber\\
&&&S_r\cap S_{r'}=\varnothing
&&\forall r\neq r'.\nonumber
\end{align}
Here, $F_r$ is the valuation induced by the chosen $k$-Accept protocol, including FA and BA as special cases. Unlike exclusive dispatch, which reduces to a maximum-weight matching problem with edge weights $p_{r,d}w_{r,d}$, non-exclusive dispatch allows several drivers to be notified for the same rider. Its value therefore depends on both the probability of receiving an acceptance and the competition among accepting drivers.

Our main result is a simple deterministic algorithm that achieves a $(1-1/e)$ approximation under every $k$-Accept protocol. We first formulate an ex-ante linear programming relaxation that upper-bounds the optimal value under all these protocols (\Cref{sec:exAnte}). Then in \Cref{sec:algMulti} we independently assign each driver to at most one rider according to the LP solution and attenuate notifications using the linear rule $a(y)=1-y/3$. The key guarantee is edgewise: under FA, every rider--driver pair is matched with probability at least $(1-1/e)$ times its LP matching marginal. Summing these guarantees establishes the approximation result for FA. Since choosing the highest-weight driver among the first $k$ acceptors yields at least the reward of choosing the first acceptor, the same algorithm achieves this approximation under every $k$-Accept protocol. We then derandomize the algorithm using conditional expectations, preserving the welfare guarantee simultaneously for all protocols in \Cref{sec:deRandomize}. 

The analysis also establishes a comparison between the protocols themselves: the optimal FA welfare is at least $(1-1/e)$ times the optimal BA welfare, and hence at least the same fraction of the optimal welfare under any $k$-Accept protocol. We show that the FA--BA bound is asymptotically tight, even for a single rider. Finally, we identify a matching integrality gap for the ex-ante LP, showing that improving the approximation guarantee relative to this relaxation is impossible. This motivates the stronger approach for BA developed in \Cref{sec:BA}.

\subsection{An Ex-Ante Linear Programming Relaxation}\label{sec:exAnte}

We formulate a linear programming relaxation using notification and matching marginals. For each rider--driver pair $(r,d)$, let $x_{r,d}$ represent the probability of notifying driver $d$ about rider $r$, and let $y_{r,d}$ represent the probability that the pair is ultimately matched. Any feasible notification policy and contention resolution protocol induce marginals satisfying
\begin{align}
\label{eq:LP}\tag{LP}
\max_{x,y\geq0}\quad
& \sum_{r\in\mathcal{R}}\sum_{d\in\mathcal{D}}
w_{r,d}y_{r,d}\\
\text{s.t.}\quad
& y_{r,d}\leq p_{r,d}x_{r,d}
&& \forall r\in\mathcal{R},\ d\in\mathcal{D}, \nonumber\\
& \sum_{r\in\mathcal{R}}x_{r,d}\leq1
&& \forall d\in\mathcal{D}, \nonumber\\
& \sum_{d\in\mathcal{D}}y_{r,d}\leq1
&& \forall r\in\mathcal{R}. \nonumber
\end{align}
The first constraint requires that a match occur only if the driver is notified and accepts. The second constraint reflects the restriction that each driver receives at most one notification, while the third reflects the restriction that each rider is matched to at most one driver. The analogous driver constraint $\sum_r y_{r,d}\leq1$ is redundant, since $y_{r,d}\leq p_{r,d}x_{r,d}\leq x_{r,d}$.

These constraints do not specify how contention among accepting drivers is resolved. Consequently, the same LP applies to every $k$-Accept protocol, including FA and BA. It is an ex-ante relaxation: every feasible policy induces feasible marginals, but a feasible LP solution need not be implementable by a notification policy. In particular, the LP does not fully capture the competition among drivers notified for the same rider.

Let $\mathrm{LP}$ denote the optimal value of \eqref{eq:LP}, and let $\mathrm{OPT}_{k\text{-Accept}}$ denote the optimal value in \eqref{eq:mainP} under the $k$-Accept protocol.

\begin{proposition}[LP upper bound and protocol comparison]
\label{prop:LP-upper-bound}
For every $k$-Accept protocol,
\[
\mathrm{LP}\geq\mathrm{OPT}_{k\text{-Accept}}.
\]
\end{proposition}
% \begin{proofof}{Proof}
%     Take $S^*$ and define $\tilde{x}_{r,d} = \Ind{d \in S_r}$ and $\tilde{y}_{r,d} = \mathbb{P}\left[r,d \text{ are matched} \right]$. We will show this is a feasible solution for \eqref{eq:LP} and the desired result will follow.
    
%     First of all, note that $\tilde{x}_{r,d},\tilde{y}_{r,d}$ are both positive and less than $1$. second, 
%     \begin{align*}
%         \tilde{y}_{r,d} = \mathbb{P}\left[r,d \text{ are matched} \right] \leq  \mathbb{P}\left[d \text{ accepts } r| r \text{ notified to } d \right]x_{r,d} = p_{r,d}x_{r,d}.
%     \end{align*}
%     Furthermore, from the fact that for all $r,r' \in \mathcal{R}$, $S_r \cap S_{r'} = \emptyset$, we get that $\Ind{d \in S_r}$ is $1$ for at most one $d$ and therefore, $\sum_{r \in \mathcal{R}} \tilde{x}_{r,d} \leq 1$ and the last inequality comes from the fact that $F_r^{\text{FA}}(\cdot)$ is the expectation of the matched weight.
% \end{proofof}
\begin{proofof}{Proof}
Fix any feasible collection of disjoint notification sets $(S_r)_{r\in\mathcal{R}}$. Define
\[
\tilde{x}_{r,d}:=\Ind{d\in S_r},
\qquad
\tilde{y}_{r,d}:=
\Pr(r\text{ is matched to }d).
\]
A pair can be matched only if the driver is notified and accepts, so
\[
\tilde{y}_{r,d}\leq p_{r,d}\tilde{x}_{r,d}.
\]
Disjointness of the notification sets implies
$\sum_r\tilde{x}_{r,d}\leq1$ for every driver $d$. Since each rider is matched to at most one driver, we also have
$\sum_d\tilde{y}_{r,d}\leq1$ for every rider $r$. Thus $(\tilde{x},\tilde{y})$ is feasible for \eqref{eq:LP}.

Its objective value equals the expected value of the notification sets:
\[
\sum_{r\in\mathcal{R}}\sum_{d\in\mathcal{D}}
w_{r,d}\tilde{y}_{r,d}
=
\sum_{r\in\mathcal{R}}F_r(S_r).
\]
Applying this construction to an optimal solution of \eqref{eq:mainP} proves the claim.
\end{proofof}

This upper bound allows us to analyze the rounding algorithm against a common benchmark for all protocols. In the following subsection, we construct feasible notification sets from an optimal LP solution and show that their expected value is at least $(1-1/e)\mathrm{LP}$ under every $k$-Accept protocol.

% This result shows that if we find an algorithm achieving an $\alpha$ approximation relative to LP, it will be also an $\alpha$ approximation relative to OPT. The algorithm is pretty simple: rounding the probabilities of acceptance independently and retaining each edge with some attenuation function.

\subsection{Independent Rounding with Attenuation}\label{sec:algMulti}

Let $(x^*,y^*)$ be an optimal solution of \eqref{eq:LP}. We construct notification sets by processing drivers independently. For each driver $d$, we first sample a rider $r$ with probability $y^*_{r,d}/p_{r,d}$, leaving the driver unassigned with the remaining probability. This distribution is well-defined because
\[
\sum_{r:p_{r,d}>0}\frac{y^*_{r,d}}{p_{r,d}}
\leq \sum_{r\in\mathcal{R}}x^*_{r,d}\leq1.
\]
Pairs with $p_{r,d}=0$ have $y^*_{r,d}=0$ and are never sampled. If rider $r$ is sampled, we retain the notification with probability
\[
a(y^*_{r,d})=1-\frac{y^*_{r,d}}3.
\]
Thus each driver receives at most one notification, and the resulting notification sets are disjoint.

The attenuation step controls contention under FA. Without attenuation, a pair with a small matching marginal can lose nearly half of its active probability to a competing pair whose marginal is close to one. For example, with two pairs having marginals $\epsilon$ and $1-\epsilon$ for the same rider, independent rounding without attenuation gives the first pair winning probability
\[
\epsilon\left(1-\frac{1-\epsilon}{2}\right)
=\frac{\epsilon(1+\epsilon)}2.
\]
The ratio to its LP marginal approaches $1/2$ as $\epsilon\to0$, preventing the desired edgewise $(1-1/e)$ guarantee. The decreasing function $a(\cdot)$ reduces notifications more strongly for pairs with larger marginals, limiting their effect on competing pairs.

\begin{algorithm}[htb]
\caption{Independent rounding with attenuation}
\label{alg:multi}
\KwIn{Weights $\{w_{r,d}\}$ and acceptance probabilities $\{p_{r,d}\}$}
\KwOut{Disjoint notification sets $(S_r)_{r\in\mathcal{R}}$}

Solve \eqref{eq:LP} to obtain an optimal solution $(x^*,y^*)$\;
Set $S_r\gets\varnothing$ for every $r\in\mathcal{R}$\;
\ForEach{$d\in\mathcal{D}$}{
    Independently sample $R_d\in\mathcal{R}\cup\{\varnothing\}$ with
    \[
    \Pr(R_d=r)=
    \begin{cases}
        y^*_{r,d}/p_{r,d}, & p_{r,d}>0,\\
        0, & p_{r,d}=0,
    \end{cases}
    \qquad
    \Pr(R_d=\varnothing)
    =1-\sum_{r:p_{r,d}>0}\frac{y^*_{r,d}}{p_{r,d}}
    \]
    \If{$R_d=r\neq\varnothing$}{
        Independently retain the notification with probability
        $1-y^*_{r,d}/3$\;
        If retained, set $S_r\gets S_r\cup\{d\}$\;
    }
}
\KwRet{$(S_r)_{r\in\mathcal{R}}$}\;
\end{algorithm}

In the remainder of this subsection, we first analyze the algorithm under FA and establish an edgewise guarantee: every pair $(r,d)$ is matched with probability at least $(1-1/e)y^*_{r,d}$. We then extend the value guarantee to every $k$-Accept protocol. The following integral inequality is the key step in the FA analysis.

\begin{lemma}
\label{lem:attenuation-integral}
Let $y_1,\ldots,y_m\in[0,1]$ and $a(y)=1-y/3$. Then
\[
\int_0^1\prod_{\ell=1}^m
\bigl(1-t y_\ell a(y_\ell)\bigr)\,dt
\geq
\int_0^1 e^{-t\sum_{\ell=1}^m y_\ell}\,dt.
\]
\end{lemma}

\begin{proposition}[Edgewise guarantee under FA]
\label{prop:FA1-1/e}
Under Algorithm~\ref{alg:multi} with $a(y)=1-y/3$ and the FA protocol, every rider--driver pair $(r,d)$ is matched with probability at least $\left(1-\frac1e\right)y^*_{r,d}$.
\end{proposition}
\begin{proofof}{Proof}
Fix a rider $r$ and driver $d$. The probability that any driver $k$ is notified about rider $r$ and accepts is
$y^*_{r,k}a(y^*_{r,k})$, independently across drivers. Hence, by the FA integral representation \eqref{eq:int-rep}, with the change of variable $t\mapsto1-t$,
\begin{align*}
\Pr(d\text{ wins }r)
&=
y^*_{r,d}a(y^*_{r,d})
\int_0^1
\prod_{k\in\mathcal{D}\setminus\{d\}}
\bigl(1-t y^*_{r,k}a(y^*_{r,k})\bigr)\,dt\\
&\geq
y^*_{r,d}a(y^*_{r,d})
\int_0^1 e^{-t\sum_{k\neq d}y^*_{r,k}}\,dt\\
&\geq
y^*_{r,d}\left(1-\frac{y^*_{r,d}}3\right)
\int_0^1 e^{-t(1-y^*_{r,d})}\,dt\\
&= y^*_{r,d}\left(1-\frac{y^*_{r,d}}3\right)\frac{1-e^{-(1-y^*_{r,d})}}{1-y^*_{r,d}}.
\end{align*}
The first inequality follows from Lemma~\ref{lem:attenuation-integral}, and the second uses the LP constraint
$\sum_k y^*_{r,k}\leq1$.

Set $s=1-y^*_{r,d}\in[0,1]$. The multiplier of $y^*_{r,d}$ in the last expression is
\[
H(s):=\frac{(2+s)(1-e^{-s})}{3s},
\]
with $H(0)=2/3$ by continuity. For $s>0$,
\[
H'(s)
=
\frac{(s^2+2s+2)e^{-s}-2}{3s^2}
\leq0,
\]
where the inequality follows from $2e^s\geq2+2s+s^2$.
Thus $H(s)\geq H(1)=1-1/e$ throughout $[0,1]$, proving the claim.
\end{proofof}

The edgewise guarantee immediately yields a value guarantee under FA. To extend it to $k$-Accept, we compare the protocols on the same notification sets, acceptance decisions, and response order. Whenever at least one driver accepts, the first acceptor is among the candidates considered by $k$-Accept. Selecting the highest-weight candidate therefore yields at least the reward obtained under FA.

\begin{theorem}[Approximation under $k$-Accept]
\label{thm:multi-approx}
For every $k$-Accept protocol, including FA and BA, Algorithm~\ref{alg:multi} with $a(y)=1-y/3$ satisfies
\[
\mathbb{E}[\mathrm{ALG}_{k\text{-Accept}}]
\geq
\left(1-\frac1e\right)\mathrm{LP}
\geq
\left(1-\frac1e\right)\mathrm{OPT}_{k\text{-Accept}}.
\]
The expectation is over the algorithm's randomization, driver acceptances, and response order.
\end{theorem}

\begin{proofof}{Proof}
Multiplying the guarantee in Proposition~\ref{prop:FA1-1/e} by the nonnegative weights and summing over all pairs gives
\[
\mathbb{E}[\mathrm{ALG}_{\mathrm{FA}}]
=
\sum_{r,d}w_{r,d}\Pr(d\text{ wins }r)
\geq
\left(1-\frac1e\right)
\sum_{r,d}w_{r,d}y^*_{r,d}
=
\left(1-\frac1e\right)\mathrm{LP}.
\]
For every realization of the notifications, acceptances, and response order, the total reward under $k$-Accept is at least that under FA. Consequently,
\[
\mathbb{E}[\mathrm{ALG}_{k\text{-Accept}}]
\geq
\mathbb{E}[\mathrm{ALG}_{\mathrm{FA}}]
\geq
\left(1-\frac1e\right)\mathrm{LP} \geq
\left(1-\frac1e\right)\mathrm{OPT}_{k\text{-Accept}}.
\]
The final inequality follows from Proposition~\ref{prop:LP-upper-bound}.
\end{proofof}
Importantly, Algorithm~\ref{alg:multi} does not depend on the contention resolution protocol. The same random notification sets therefore achieve the $(1-1/e)$ approximation guarantee in expectation under all $k$-Accept protocol; no change to the LP solution or the rounding procedure is needed when switching protocols. This observation also yields a comparison between their optimal value values.

\begin{corollary}[Comparison of optimal protocol value]
\label{cor:FA-BA-comparison}
For every $k$-Accept protocol,
\[
\left(1-\frac1e\right)\mathrm{OPT}_{\mathrm{BA}}
\leq
\mathrm{OPT}_{\mathrm{FA}}
\leq
\mathrm{OPT}_{k\text{-Accept}}
\leq
\mathrm{OPT}_{\mathrm{BA}}.
\]
\end{corollary}

The factor in Corollary~\ref{cor:FA-BA-comparison} is tight, even with a single rider and a common acceptance probability for all drivers. The following example captures the trade-off between a rare, high-value match and a reliable, low-value fallback. Under BA, the platform can obtain the fallback without sacrificing the high-value match whenever it is available. Under FA, additional fallback drivers increase competition against the high-value driver.

\begin{example}[Tight FA--BA comparison]
\label{ex:tight-FA-BA}
Consider one rider and $n^2+1$ drivers, where $n\geq2$. Driver $0$ has weight $w_0=1$, while the remaining $n^2$ drivers have weight
\[
w_d=\frac{e-2}{n},
\qquad d=1,\ldots,n^2.
\]
Every driver accepts independently with probability $1/n$. Then
\[
\lim_{n\to\infty}
\frac{\mathrm{OPT}_{\mathrm{FA}}}
     {\mathrm{OPT}_{\mathrm{BA}}}
=1-\frac1e.
\]
\end{example}

\begin{proofof}{Proof}
Under BA, it is optimal to notify all drivers, hence
\[
\mathrm{OPT}_{\mathrm{BA}}
=
\frac1n+
\left(1-\frac1n\right)\frac{e-2}{n}
\left[1-\left(1-\frac1n\right)^{n^2}\right],
\]
so
\[
\lim_{n\to\infty}n\,\mathrm{OPT}_{\mathrm{BA}}=e-1.
\]

Under FA, an optimal notification set should include driver $0$, since it has maximum weight (\Cref{lem:max-weight}). All remaining drivers are identical, so the decision reduces to choosing their number. Let $F(k)$ denote the FA value from notifying driver $0$ and $k$ fallback drivers, where $0\leq k\leq n^2$. If
$K\sim\operatorname{Bin}(k,1/n)$ is the number of accepting fallback drivers, conditioning on driver $0$'s acceptance gives
\[
F(k)
=
\left(1-\frac1n\right)\frac{e-2}{n}\Pr(K\geq1)
+
\frac1n\,
\mathbb{E}\left[\frac{1+\frac{e-2}{n}K}{1+K}\right].
\]
% Using
% \[
% \mathbb{E}\left[\frac1{1+K}\right]
% =
% \frac{n}{k+1}
% \left[1-\left(1-\frac1n\right)^{k+1}\right],
% \]
% we obtain the exact expression
% \[
% F(k)
% =
% \left(
% \frac{e-2}{n}
% +\frac{1-\frac{e-2}{n}}{k+1}
% \right)
% \left[1-\left(1-\frac1n\right)^{k+1}\right].
% \]
For any sequence $k_n$ with $k_n/n\to\lambda<\infty$, the distribution $K\sim\operatorname{Bin}(k,1/n)$ converges to Poisson with rate $\lambda$:
\[
nF(k_n)\longrightarrow
g(\lambda):=
(e-2)(1-e^{-\lambda})
+\frac{1-e^{-\lambda}}{\lambda},
\]
where $g(0)=1$ by continuity. We next show that $g$ is maximized at $\lambda=1$. For $\lambda>0$,
\[
g'(\lambda)
=
\frac{e^{-\lambda}}{\lambda^2}
\left[1+\lambda+(e-2)\lambda^2-e^\lambda\right].
\]
The function
\[
\frac{e^\lambda-1-\lambda}{\lambda^2}
=
\sum_{j=0}^{\infty}\frac{\lambda^j}{(j+2)!}
\]
is strictly increasing for $\lambda>0$ and equals $e-2$ at $\lambda=1$. Thus $g'(\lambda)>0$ for $\lambda<1$ and $g'(\lambda)<0$ for $\lambda>1$. Consequently,
\[
\max_{\lambda\geq0}g(\lambda)
=g(1)
=(e-1)\left(1-\frac1e\right).
\]

If $k_n/n\to\infty$, the exact expression above instead gives
\[
nF(k_n)\longrightarrow e-2<g(1).
\]
These cases establish
\[
\lim_{n\to\infty}n\,\mathrm{OPT}_{\mathrm{FA}}
=(e-1)\left(1-\frac1e\right).
\]
Dividing by the corresponding limit for BA proves the claim.
\end{proofof}

Finally, it is worth noting that the $(1-1/e)$ guarantee relative to \eqref{eq:LP} is asymptotically tight. Consider one rider and $n$ drivers, each with weight $1$ and acceptance probability $1/n$. The LP has value $1$, attained by $x_d=1$ and $y_d=1/n$. Under every $k$-Accept protocol, however, the optimal welfare is the probability that at least one driver accepts:
\[
\mathrm{OPT}_{k\text{-Accept}}
=1-\left(1-\frac1n\right)^n
\longrightarrow 1-\frac1e.
\]
Thus, to overcome the $1-1/e$ barrier through LP-based rounding, we need a tighter relaxation. As a first step in this direction, \Cref{sec:BA} considers the configuration LP for BA and uses tools from submodular value maximization to obtain an approximation factor strictly greater than $1-1/e$.

\subsection{Derandomization}
\label{sec:deRandomize}

The randomization in Algorithm~\ref{alg:multi} can be removed without weakening the approximation guarantee. We process drivers in a fixed order one-by-one and choose each notification to maximize the value under FA, keeping the remaining notification decisions randomized. The idea is that, for each driver whose notification is still random, we select the rider with the largest positive marginal gain relative to leaving the driver unnotified. The resulting notification sets are deterministic and satisfy the same guarantee under every $k$-Accept protocol. 

To implement this procedure, with a slight abuse of notation, we evaluate the FA value under effective acceptance probabilities $\boldsymbol{u}$. For each rider $r$, define
\begin{align*}
F_r^{\text{FA}}(\boldsymbol{u}_r)\coloneqq
\sum_{d\in\mathcal{D}}w_{r,d}
\,u_{r,d}
\int_0^1
\prod_{\ell\in\mathcal{D}\setminus\{d\}}
(1-u_{r,\ell}t)\,dt.
\end{align*}
This is the integral representation of \Cref{lem:int-rep} with the acceptance probabilities replaced by $\boldsymbol{u}_r$, written after the change of variable $t\mapsto 1-t$. Here, $u_{r,d}$ is the probability that driver $d$ is \emph{notified} about rider $r$ and accepts, equal to $p_{r,d}x_{r,d}$ in \eqref{eq:LP}. In particular, if driver $d$ is notified about rider $r$, we have $u_{r,d}=p_{r,d}$ for that rider and zero for every other rider. If it is left unnotified, the conditional effective acceptance probability becomes zero for every rider. Initially, we start from the randomized solution and set $u_{r,d}=y^*_{r,d}a(y^*_{r,d})$; we denote this initial vector by $\boldsymbol{u}_r^{\mathrm{initial}}$. 

For each driver $d$, we first evaluate the outcome of leaving it unnotified. We then compute the gain from assigning it to each rider and select a rider with the largest positive gain, if one exists. Let $\boldsymbol{u}_r^{\,d\leftarrow p_{r,d}}$ denote the vector obtained from $\boldsymbol{u}_r$ by replacing coordinate $d$ with $p_{r,d}$.

\begin{algorithm}[htb]
\caption{Deterministic rounding by conditional expectations}
\label{alg:derandomized}
\KwIn{Weights $\{w_{r,d}\}$ and acceptance probabilities $\{p_{r,d}\}$}
\KwOut{Disjoint notification sets $(S_r)_{r\in\mathcal{R}}$}

Solve \eqref{eq:LP} to obtain an optimal solution $(x^*,y^*)$\;
Set $S_r\gets\varnothing$ for every $r\in\mathcal{R}$\;
Set $u_{r,d}\gets y^*_{r,d}(1-y^*_{r,d}/3)$
for every $(r,d)$\;

\ForEach{$d\in\mathcal{D}$ in a fixed order}{
    Set $u_{r,d}\gets0$ for every $r\in\mathcal{R}$\;
    \ForEach{$r\in\mathcal{R}$}{
        Compute
        \[
        \Delta_{r,d}
        \gets
        F_r^{\text{FA}}(\boldsymbol{u}_r^{\,d\leftarrow p_{r,d}})
        -F_r^{\text{FA}}(\boldsymbol{u}_r)
        \]
    }
    Choose $r^*\in\arg\max_r\Delta_{r,d}$,
    breaking ties deterministically\;
    \If{$\Delta_{r^*,d}>0$}{
        Set $S_{r^*}\gets S_{r^*}\cup\{d\}$\;
        Set $u_{r^*,d}\gets p_{r^*,d}$\;
    }
}
\KwRet{$(S_r)_{r\in\mathcal{R}}$}\;
\end{algorithm}

\begin{theorem}[Deterministic approximation]
\label{thm:derandomized}
Algorithm~\ref{alg:derandomized} runs in polynomial time and returns deterministic notification sets satisfying, simultaneously for every $k$-Accept protocol,
\[
\sum_{r\in\mathcal{R}}F_r^{k\text{-Accept}}(S_r)
\geq
\left(1-\frac1e\right)\mathrm{LP}
\geq
\left(1-\frac1e\right)\mathrm{OPT}_{k\text{-Accept}}.
\]
\end{theorem}

\begin{proofof}{Proof}
Initially, $\sum_r F_r^{\text{FA}}(\boldsymbol{u}_r)$ equals the expected FA value of Algorithm~\ref{alg:multi}. By Theorem~\ref{thm:multi-approx},
\[
\sum_rF_r^{\text{FA}}(\boldsymbol{u}_r^{\mathrm{initial}})
\geq
\left(1-\frac1e\right)\mathrm{LP}.
\]
We show that fixing each driver's notification cannot decrease this value.

Fix the current driver $d$, and let $C_0$ denote the conditional expected value when this driver is left unnotified. Assigning it to rider $r$ instead gives value $C_0+\Delta_{r,d}$. Consequently, the conditional expected value before processing $d$ is
\[
C_0+
\sum_{r:p_{r,d}>0}
\frac{y^*_{r,d}a(y^*_{r,d})}{p_{r,d}}
\Delta_{r,d}.
\]
Therefore, since the coefficients $\frac{y^*_{r,d}a(y^*_{r,d})}{p_{r,d}}$ are nonnegative and satisfy $\sum_{r:p_{r,d}>0}\frac{y^*_{r,d}a(y^*_{r,d})}{p_{r,d}}\leq\sum_{r:p_{r,d}>0}\frac{y^*_{r,d}}{p_{r,d}}\leq 1$ (see \Cref{sec:algMulti}), the expression above is at most
\[
C_0+\max\left\{0,\max_r\Delta_{r,d}\right\}.
\]
This is precisely the value obtained by the algorithm's choice. Thus the conditional expected FA value never decreases. Once all notifications have been fixed, we obtain
\[
\sum_{r\in\mathcal{R}}F_r^{k\text{-Accept}}(S_r)
\geq \sum_{r\in\mathcal{R}}F_r^{\text{FA}}(S_r)
\geq
\left(1-\frac1e\right)\mathrm{LP}.
\]
For these same notification sets, every $k$-Accept protocol yields at least the FA reward under the same acceptance decisions and response order. Proposition~\ref{prop:LP-upper-bound} then gives the claimed guarantee.
\end{proofof}

The derandomization preserves total expected value, rather than the edgewise matching guarantees of the randomized algorithm. The notification sets themselves are deterministic; the value values still average over driver acceptances and response order. A single collection of sets therefore provides the approximation guarantee under every $k$-Accept protocol, with no additional approximation loss.

\section{Best Acceptance NED Notification Problem: Improved Guarantees}
\label{sec:BA}
In this section, we study welfare maximization under BA valuations defined in Definition~\ref{def:BA}. From a marketplace-design perspective, BA corresponds to a ``wait-and-choose'' policy: the platform can broadcast more broadly to improve match quality, but must accept a confirmation delay to collect responses. As mentioned in \Cref{sec:BA-def}, the single-rider problem is trivial due to monotonicity of BA valuation, and therefore we focus on the general multi-rider setting.

We begin by examining a practical yet tractable special case in which acceptance probabilities are homogeneous across drivers for each rider in \Cref{sec:fixedP}. We then address the general case with heterogeneous \FEedit{rejection} probabilities in \Cref{sec:general_case} and present approximation algorithms.

\subsection{Rider-Specific Acceptance Probabilities: an Exact Polynomial-time Solution}
\label{sec:fixedP}
Consider the case where the acceptance probabilities are identical across drivers for each rider, denoted by $p_{r,d}=p_r\in(0,1)$ for all $r\in\RiderSet, d\in\DriverSet$. This setting is operationally meaningful, as it captures the practical scenario in which the platform makes notification decisions without access to individual \FEedit{rejection} probabilities and only uses an average quantity $p_r$ for each rider.\footnote{Recently, there has been more emphasis in ride-sharing platforms on algorithms/mechanisms that are not penalizing drivers for their behavior as gig workers, and \FEedit{rejection}-unaware decision making certainly belongs to this category.} The common-probability setting $p_r=p$ for all riders is a special case.

Fix a rider $r$ and a notified set $S$. If we sort drivers in $S$ in non-increasing order of $\Weight{r}{d}$ and denote these scores by
$\Weight{r}{1}\ge \Weight{r}{2}\ge\cdots$, then
\begin{align}
\Fi{r}{S} = \sum_{\ell=1}^{|S|} p_r(1-p_r)^{\ell-1}\,\Weight{r}{\ell}.
\label{eq:BA-homog-sorted}
\end{align}
In words, since the acceptance probability $p_r$ is constant across drivers for a fixed rider $r$, the probability that this rider successfully matches with the $\ell$-th driver in their sorted list of weights depends solely on the index $\ell$, regardless of which specific drivers occupy positions $1$ through $\ell-1$. This structure yields an exact polynomial-time solution for the maximum welfare problem through a reduction to maximum-weight bipartite matching. Formally, we show the following proposition.

\begin{proposition}[Rider-specific-$p_r$ BA reduces to maximum weight matching]
\label{prop:BA-homogeneous-matching}
Suppose $\AccProb{r}{d}=p_r\in(0,1)$ for all $(r,d)$. Consider a bipartite graph with driver nodes on the left, and on the right create $\NumDrivers$ \emph{slots} for each rider, indexed by $(r,\ell)$ for $\ell\in\{1,\dots,\NumDrivers\}$. For each pair $(r,d)$ and slot index $\ell$, add an edge between driver $d$ and slot $(r,\ell)$ with weight $p_r(1-p_r)^{\ell-1}\Weight{r}{d}$. Then the BA welfare maximization problem is solvable exactly in polynomial time by computing a maximum-weight matching in this expanded bipartite graph. Equivalently, the following LP admits an integral optimal solution:
\begin{align*}
    \underset{x_{r,d,\ell}\geq 0}{\max} \quad &
    \sum_{r\in\RiderSet}\sum_{d\in\DriverSet}
    \sum_{\ell=1}^{\NumDrivers}
    p_r(1-p_r)^{\ell-1}\,\Weight{r}{d}\,x_{r,d,\ell} \\
    \text{s.t.} \quad &
    \sum_{r\in\RiderSet}\sum_{\ell=1}^{\NumDrivers}
    x_{r,d,\ell}\le 1
    \quad && \forall d\in\DriverSet, \\
    & \sum_{d\in\DriverSet}x_{r,d,\ell}\le 1
    \quad && \forall r\in\RiderSet,\;
    \forall \ell\in\{1,\dots,\NumDrivers\}.
\end{align*}
\end{proposition}

\begin{proofof}{Proof}
First, note that the constraint matrix is that of a bipartite matching (drivers vs. rider-slots), so the LP is totally unimodular and admits an integral optimal solution. Now, given a feasible notification allocation $(S_1,\dots,S_{\NumRiders})$, sort each $S_r$ by $\Weight{r}{d}$ in non-increasing order and assign its $\ell$-th driver to slot $(r,\ell)$. This produces a feasible matching with total weight equal to $\sum_r \Fi{r}{S_r}$ by \eqref{eq:BA-homog-sorted}.

Conversely, given any integral maximum-weight matching $x^*_{r,d,\ell}$, define $S_r$ to be the set of drivers matched to slots of rider $r$. Within a fixed rider $r$, if two matched drivers $d$ and $d'$ are assigned to slots $\ell<\ell'$ but have $\Weight{r}{d}<\Weight{r}{d'}$, swapping them weakly increases the matching objective because
$p_r(1-p_r)^{\ell-1}\ge p_r(1-p_r)^{\ell'-1}$.
Thus, there exists an integral maximum-weight matching in which the matched drivers for each rider are sorted by score across slots.

For such a matching, the $\ell$-th driver in $S_r$, sorted in non-increasing order of weights, is matched to a slot $(r,\ell^\dagger)$ where $\ell^\dagger\geq\ell$. If $\ell^\dagger>\ell$ for some driver, we can shift the assignments of drivers in $S_r$ towards lower-index slots. This weakly increases the objective function, since $p_r(1-p_r)^{\ell-1}$ is decreasing in $\ell$ for $p_r\in(0,1)$ and the weights are nonnegative. Therefore, there always exists an integral optimal matching $x^*_{r,d,\ell}$ such that, for each $r\in\RiderSet$, the $\ell$-th driver in $S_r$ is matched to slot $(r,\ell)$. For such a matching, the induced sets $(S_1,\dots,S_{\NumRiders})$ achieve a BA welfare that is exactly equal to the total weight of matching $x^*_{r,d,\ell}$, i.e., the objective of the LP in the statement of Proposition~\ref{prop:BA-homogeneous-matching}, again by \eqref{eq:BA-homog-sorted}.
\end{proofof}

\subsection{General Case: Reduction to Submodular Welfare Maximization}
\label{sec:general_case}
In the general setting where acceptance probabilities $p_{r,d}$ vary across drivers and riders, the exact matching formulation described in Section~\ref{sec:fixedP} does not apply, as the probability of reaching rank $\ell$ depends on the specific identities of the preceding drivers. However, a key observation is that BA valuation functions are monotone and submodular. Consequently, this problem is a special instance of the Submodular Welfare Maximization (SWM) problem~\citep{vondrak2008optimal}.

It is a well-established result that the general SWM problem admits a $(1 - 1/e)$-approximation algorithm in the \emph{value-oracle} model (e.g., via the continuous greedy algorithm and matroid rounding methods such as pipage rounding or swap rounding~\citep{vondrak2008optimal}). This provides a baseline theoretical guarantee for BA welfare maximization problem. 

Moreover \cite{feige2010submodular} show that the $(1 - 1/e)$ gap can be broken and obtain a polynomial-time algorithm (based on solving configuration LP combined with a specific correlated rounding algorithm) with an approximation factor $(1 - 1/e + c)$  for a small universal constant $c>0$ with a \emph{demand oracle}:
recall that the demand oracle, given non-negative prices (or dual costs) $\{\lambda_d\}_{d\in\DriverSet}$ on drivers, computes an (approximately) optimal solution to
\begin{equation}
\max_{S\subseteq \DriverSet}\left(\Fi{r}{S}-\sum_{d\in S}\lambda_d\right).
\label{eq:BA-demand}
\end{equation}
While $\Fi{r}{S}-\sum_{d\in S}\lambda_d$ is not monotone because of the cost term, the special form of $F_r(\cdot)$ under BA allows a near-optimal solution via knapsack-style dynamic programming after discretizing the price space. Formally, we show the following proposition, with proof postponed to Appendix~\ref{apx:proof-DP-demand-oracle}.

\begin{proposition}[Additive-$\epsilon$ demand oracle for BA]
\label{prop:BA-demand-DP}
Fix a rider $r$ and the BA valuation function $F_r(\cdot)$ defined in~\eqref{eq:Fi-BA-def}.
For every $\epsilon>0$, there exists an additive-$\epsilon$ demand oracle that runs in time
$\textrm{poly}\!\left(n,\frac{1}{\epsilon}\right)=\mathcal{O}(n^3/\epsilon)$ and, given any nonnegative price vector
$\{\lambda_d\}_{d\in\DriverSet}$, returns a set $\hat S\subseteq \DriverSet$ such that
\[
F_r(\hat S)-\sum_{d\in \hat S}\lambda_d
\;\ge\;
\max_{S\subseteq \DriverSet}\Bigl(F_r(S)-\sum_{d\in S}\lambda_d\Bigr) \;-\;\epsilon.
\]
\end{proposition}

Putting all the pieces together, we can prove the following improved approximation guarantee. 
\begin{theorem}[Beating $1-1/e$ for BA welfare maximization]\label{thm:beating1-1/e}
 Invoking the (randomized) approximation algorithm in \cite{feige2010submodular},  with access to the additive-$\epsilon$ demand oracle in \Cref{prop:BA-demand-DP} as the demand oracle, results in a randomized algorithm for BA welfare maximization problem that runs in $\textrm{poly}(n,m,\frac{1}{\epsilon})$, and returns notification sets $(\hat{S}_1,\ldots,\hat{S}_m)$ such that:
 $$
 \E\left[\sum_{r\in \RiderSet}F_r(\hat{S}_r)\right]\geq \left(1-1/e+c\right)\textrm{OPT}-\epsilon~,
 $$
 where $\textrm{OPT}$ is the optimal objective values of BA welfare maximization problem. 
\end{theorem}
\begin{proofof}{Proof sketch}
The algorithm in \cite{feige2010submodular} starts by solving the configuration LP for the underlying SWM problem. The demand oracle is essentially a separation oracle for the dual LP, and therefore with access to an exact separation oracle and running the ellipsoid algorithm, the primal LP can be solved exactly. With approximate separation oracle, we use an approximate variant of Grotschel--Lovasz--Schrijver \citep{grotschel1981ellipsoid} to solve the configuration LP up to an additive error $\epsilon$ in polynomial-time. We finally use the correlated rounding algorithm in \cite{feige2010submodular} that is oblivious to the choice of the feasible fractional point in the configuration LP. 
\end{proofof}

\section{Numerical Experiment}
\label{sec:numerical}
In this section, we present numerical experiments to evaluate the empirical performance of our proposed algorithm (Algorithms~\ref{alg:multi} and \ref{alg:derandomized}) for the Notification Set Selection problem. We assess the algorithms under both the First Accept (FA) and Best Accept (BA) protocols, comparing our proposed approaches against standard heuristics and the exact optimal solutions.

\subsection{Experimental Setup and Data Generation}
We conduct our evaluations on two types of datasets: synthetically generated instances and a real-world dataset provided by Lyft. For the synthetic data, we simulate a matching market with $m$ ride requests and $n$ available drivers. To generate an instance, we consider all $m \times n$ possible ride-driver pairs. For each pair, the match value (score) and the acceptance probability are drawn independently and uniformly at random from the interval $[0, 1]$.

The real instances are single-cycle snapshots of the Lyft market in \cite{LyftII2026companion_NonBlind}. We run the exclusive-dispatch simulation on the Lyft data 50 times, each with a fresh draw of driver types and ride multipliers, and record the pool of waiting riders and idle drivers at each of the 500 cycles, giving 25,000 snapshots. Similar to the simulations in that paper, the score of a rider--driver pair is $1/(1+\text{distance})$ and acceptance probabilities follow the driver types and ride multipliers, and only pairs with score above 0.5 are retained.

We evaluate and compare the following algorithms:
\begin{itemize}
    \item \textbf{ED:} A baseline heuristic representing an Exclusive Dispatching rule.
    \item \textbf{Greedy (FA \& BA):} The standard marginal-value greedy heuristics applied to the FA and BA objectives, respectively. In each iteration, the algorithm evaluates all drivers in a random order and adds the driver to the notification set of the ride that yields the maximum marginal increase in the objective value.
    \item \textbf{Approximation Algorithms (FA \& BA):} The primary approximation algorithms proposed in \Cref{sec:multi-ride}. Specifically, we implement Algorithm \ref{alg:multi} to get a randomized solution and then derandomize it using Algorithm \ref{alg:derandomized}.
    \item \textbf{OPT (FA \& BA):} The exact optimal solutions for both protocols, computed to serve as the ground truth for evaluating empirical approximation ratios.
\end{itemize}

\subsection{Computational Tractability}

Computing the exact optimum ($\textrm{OPT}$) under either FA or BA quickly becomes intractable as the instance size grows. As shown in Table~\ref{tab:running_times}, an instance with $m=3$ riders and $n=9$ drivers can be solved in under $0.03$ seconds, whereas an instance with $m=6$ riders and $n=18$ drivers requires more than $30$ seconds.\footnote{We implemented the exact optimization procedure with careful attention to computational efficiency.} Memory also becomes a limiting factor: on our computing environment, instances \revcolor{of size} $(m,n)=(7,21)$ \revcolor{or larger} exceeded the available memory and \revcolor{could not} be solved at all.

\begin{table}[htb]
\centering
\begin{tabular}{@{} l r r r r r r r r @{}}
\toprule
& \multicolumn{8}{c}{\textbf{Market size $(m, n)$}} \\
\cmidrule(l){2-9}
\textbf{Algorithm} & \textbf{(3,9)} & \textbf{(4,12)} & \textbf{(5,12)} & \textbf{(4,15)} & \textbf{(5,15)} & \textbf{(6,18)} & \textbf{(7,18)} & \textbf{(7,21)} \\
\midrule
\textsc{Ed}     & 1.25  & 1.57   & 1.99   & 2.08       & 2.60       & 4.63        & 4.64        & 5.57  \\
\textsc{Greedy} & 1.73  & 2.65   & 2.67   & 4.03       & 3.75       & 5.01        & 5.01        & 6.90  \\
\textsc{Approx} & 7.05  & 9.55   & 10.35  & 14.87      & 15.11      & 22.26       & 22.58       & 29.31 \\
\textsc{Opt} & 27.33 & 232.62 & 306.49 & 2{,}238.44 & 2{,}693.80 & 31{,}715.90 & 38{,}672.48 & $\infty$    \\
\bottomrule
\end{tabular}
\caption{Average running time (milliseconds) under the BA protocol, over 10 random instances
of each size, with $m$ riders and $n$ drivers. Every rider--driver pair is eligible, and
scores and acceptance probabilities are drawn uniformly from $[0,1]$. OPT is not reported
for $(7,21)$ because these instances exceeded the memory of our local computer.}
\label{tab:running_times}
\end{table}

\subsection{Results and Discussion}
We visualize the distribution of the performance ratios of our algorithms and baselines relative to the optimal solutions. The empirical approximation ratios are plotted as histograms to capture the variance and average performance across multiple randomly generated instances.
\begin{figure}[htb]
    \centering
\includegraphics[width=0.98\linewidth]{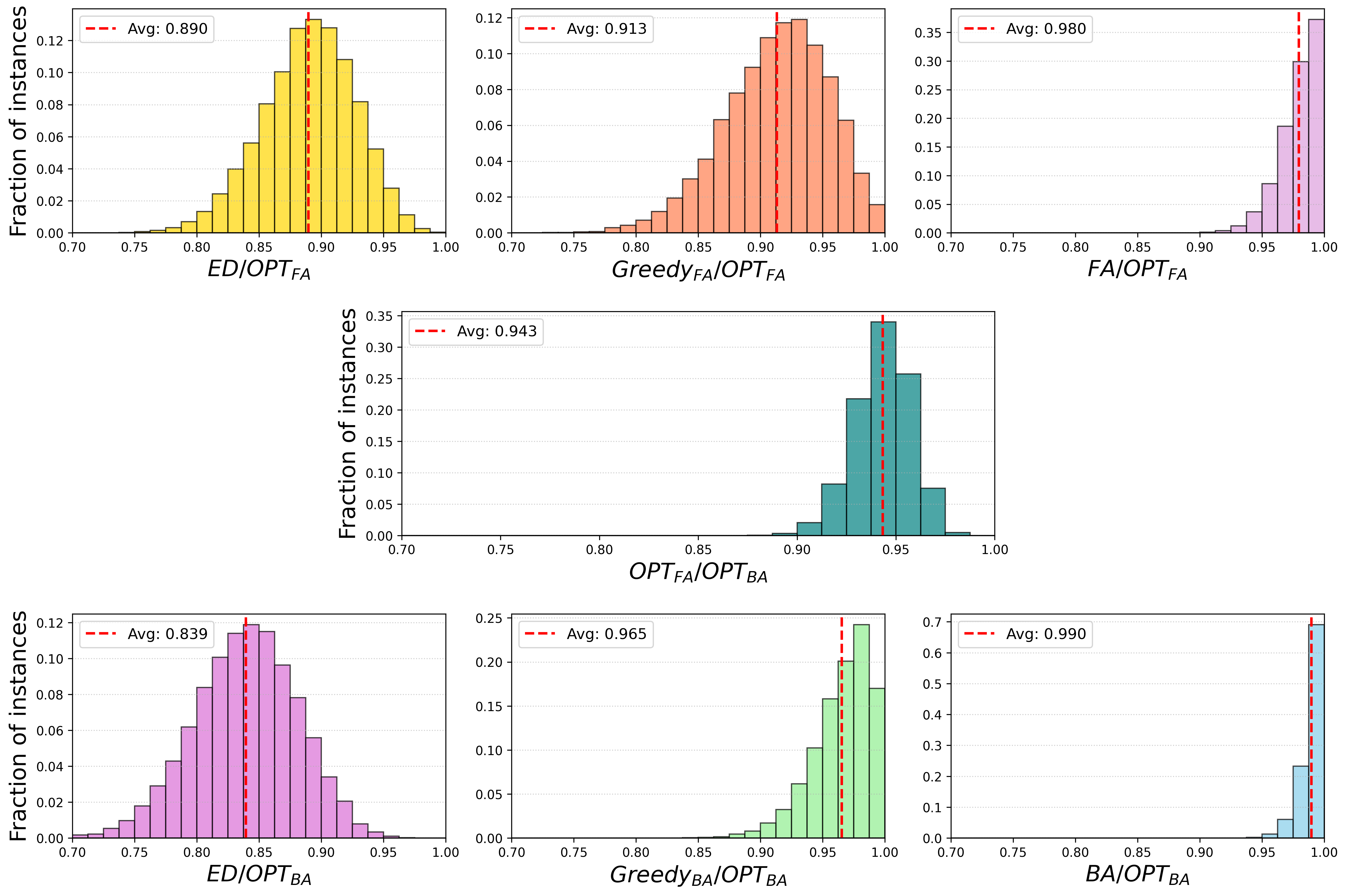} 
    \caption{Distribution of performance ratios for different algorithms evaluated on Synthetic Data ($m=5, n=15$).}
    \label{fig:synData}
\end{figure}
Figure \ref{fig:synData} illustrates the performance distributions on 25,000 synthetic instances. Our proposed FA and BA algorithms consistently achieve near-optimal performance, with average approximation ratios of 0.980 and 0.990, respectively. They clearly outperform both the standard Greedy heuristics (averaging 0.913 and 0.965) and the ED baseline (0.890 and 0.839), demonstrating the robustness of our approach. Furthermore, it is important to note that the empirical minimum performance across all simulated instances far exceeds our theoretical worst-case guarantees; while the algorithms are theoretically bounded at $1-1/e$, the actual observed minimums remain strictly above 0.9.

\begin{figure}[htb]
    \centering
\includegraphics[width=0.98\linewidth]{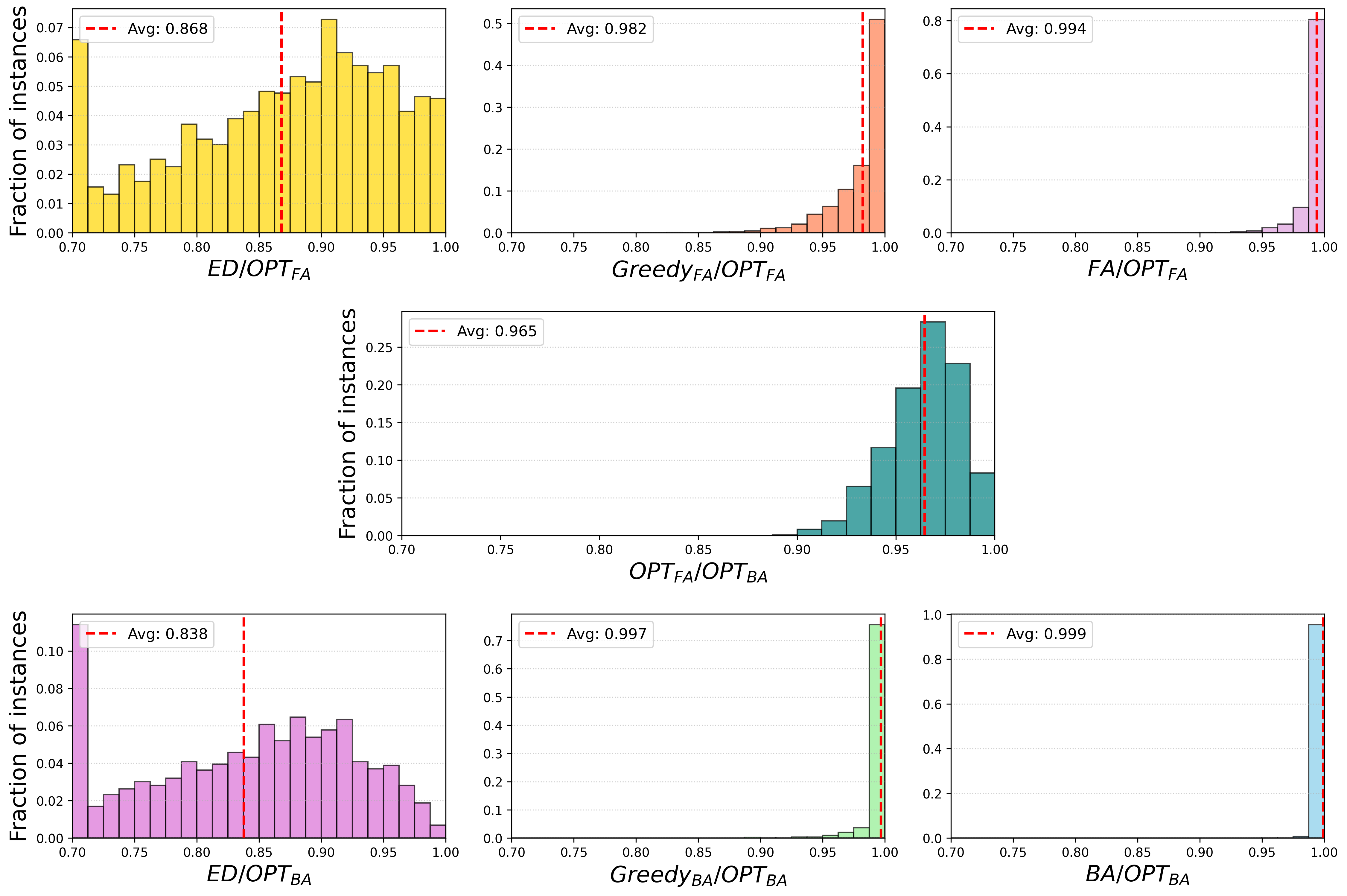}
    \caption{Distribution of performance ratios against OPT for different algorithms evaluated on the Lyft Data.}
    \label{fig:realData}
\end{figure}

Figure \ref{fig:realData} demonstrates the algorithms' performance distributions on the Lyft real-world dataset. Remarkably, on real-world data, our proposed algorithms are virtually indistinguishable from the exact optimum, achieving tight distributions with average performance ratios of 0.994 for FA and 0.999 for BA. They also outperform both the standard Greedy heuristics (averaging 0.982 and 0.997) and the single-dispatch ED baseline (0.868 and 0.838). Similar to the synthetic experiments, the empirical minimum performance far exceeds our theoretical worst-case guarantees of FA and BA. The primary reason for this reduced gap between the approximation algorithms and the optimal solutions is the underlying graph structure: while real-world instances might contain more nodes, the valid edges are much sparser either because of platform filters or geographic proximity constraints. This inherent sparsity naturally simplifies the combinatorial complexity, allowing our algorithms to consistently find near-optimal configurations.

Additionally, the gap between ED and OPT deserves a comment. Exclusive dispatch can use at
most one driver per rider, whereas OPT can notify several, so the gap is driven by how many
drivers are available per rider. In the synthetic instances this ratio is fixed, whereas in
the real snapshots it varies widely. On the real data, the gap widens steadily as the number
of drivers per rider grows: ED is close to optimal in snapshots with few drivers per rider,
and falls well short of it where drivers are plentiful. The benefit of non-exclusive
notifications is therefore largest in markets with many drivers per rider.

Finally, this single-cycle gap should not be read as the market-level gain from non-exclusive notifications: within one cycle additional notifications are free, whereas in the dynamic system of \cite{LyftII2026companion_NonBlind} a notified driver is unavailable to other riders until responding. The experiments here measure how well the algorithms solve the per-cycle optimization problem; the long-run comparison is the subject of that paper.

\section{Conclusion \& Open Problems}
\label{sec:conclusion}
In this work, as part of our collaboration with Lyft---resulting in a two-part paper, including this paper and our companion paper~\ifblind\cite{LyftII2026companion_Blind}\else\cite{LyftII2026companion_NonBlind}\fi---we model the non-exclusive dispatch process using two natural contention resolution mechanisms: First Acceptance (FA) and Best Acceptance (BA). Our analysis reveals distinct, yet closely related, computational landscapes for these two models.

Our study of the FA model, which can also be viewed as a distinct probabilistic choice model, yields several constructive algorithmic results. We provide a PTAS for the single-rider problem and a deterministic polynomial-time $(1-1/e)$-approximation algorithm for the general multi-rider setting. The same $(1-1/e)$ guarantee also applies under BA. At the same time, we show that welfare maximization is NP-hard under both FA and BA, including the single-rider FA problem, thereby ruling out a polynomial time exact solution unless $\mathrm{P}=\mathrm{NP}$.

For the BA valuation class, we establish that the welfare function is monotone and submodular. Beyond the resulting $(1-1/e)$ approximation guarantee, we exploit additional structure of the BA model to obtain an approximation guarantee that surpasses the classical $(1-1/e)$ barrier. We also identify a significant tractable special case: when acceptance probabilities are homogeneous, the BA problem admits an exact polynomial-time solution via a linear programming formulation.

While our focus here is single-cycle optimization, the system-level impact of non-exclusive dispatch depends on how these per-cycle decisions interact across time. For example, broadcasting can temporarily reserve multiple drivers for the same request and thereby affect future market thickness. Our companion paper~\ifblind\cite{LyftII2026companion_Blind}\else\cite{LyftII2026companion_NonBlind}\fi ~complements the present algorithmic analysis by studying these long-run effects through simulations and a stylized marketplace model.

\smallskip
\xhdr{Future directions \& open problems.} A natural first step is to narrow the gap between our hardness results and the best known approximation guarantees for both FA and BA. Beyond this, it would be interesting to extend the model to settings with correlated driver acceptance decisions, rather than independent acceptances. Another direction is to incorporate a cost for each notified driver, capturing the operational or engagement cost of sending additional notifications and creating an explicit trade-off between match quality and notification intensity. These extensions would bring the model closer to practical dispatch settings while raising new algorithmic and structural questions.

% \begin{APPENDICES}
% \input{tex/appendix}
% \end{APPENDICES}

% \newpage
% \section{draft}
% \RN{This is a standalone version of all the results that we probably want to include in the paper related to FA (I assume a lot of proofs go to the appendix). I suggest check all of our notes and see if there are examples or insights, or even subsidiary results, that I have not included and we may want to include in the paper.}
% \input{tex-EC/draft.tex}

% \section{Numerical Simulations}
% \label{sec:numerical}
% \input{tex-EC/numerical.tex}

\newpage

\newcommand{\newblock}{}
\setlength{\bibsep}{0.0pt}
\bibliographystyle{plainnat}
{\footnotesize
\bibliography{refs}}

% \newpage
\renewcommand{\theHchapter}{A\arabic{chapter}}
\renewcommand{\theHsection}{A\arabic{section}}

\newpage
\clearpage
%\SingleSpaced
\normalsize
\pagestyle{ECheadings}%
\ECHowEquations
\ECHowSections
\setcounter{figure}{0}%
\renewcommand\thefigure{EC.\@arabic\c@figure}%
\setcounter{table}{0}%
\renewcommand\thetable{EC.\@arabic\c@table}%
\setcounter{page}{1}\def\thepage{ec\arabic{page}}%
\hspace*{1em}

\ECDisclaimer

% \section{Further Related Work}
% \label{sec:related}
% \input{tex/related-work.tex}
\section{Missing Proofs \& Technical Details }
\subsection{Proof of Theorem~\ref{thm:hardness}}
\label{appendix:hardness}
We start the analysis with the following simple observation.
\begin{lemma}[FA/BA valuation with unit weights]
\label{lem:unitweights}
Fix any rider ${r}\in\RiderSet$ and assume ${\Weight{r}{d}}=1$ for all ${d}\in\DriverSet$. Then for every $S\subseteq\DriverSet$, under either contention resolution protocol,
\[
{\Fi{r}{S}
=
\Prob\!\left(\sum_{d\in S}\AccRV{r}{d} > 0\right)
=
1-\prod_{d\in S}\bigl(1-\AccProb{r}{d}\bigr).}
\]
\end{lemma}

\begin{proofof}{Proof}
If ${\Weight{r}{d}}\equiv 1$, then there is no difference between FA and BA. Moreover, for any realization ${(\AccRV{r}{d})_{d\in S}}$ we have
\[
{\frac{\sum_{d\in S}\Weight{r}{d}\AccRV{r}{d}}{\sum_{d\in S}\AccRV{r}{d}}
=
\frac{\sum_{d\in S}\AccRV{r}{d}}{\sum_{d\in S}\AccRV{r}{d}}
=
\begin{cases}
1, & \text{if }\sum_{d\in S}\AccRV{r}{d} > 0,\\
0, & \text{if }\sum_{d\in S}\AccRV{r}{d} = 0,
\end{cases}}
\]
where the second case uses the convention that the ratio is defined to be $0$
when the denominator is $0$. Hence, taking expectations yields ${\Fi{r}{S}=\Prob\!\left(\sum_{d\in S}\AccRV{r}{d} > 0\right)=1-\prod_{d\in S}\bigl(1-\AccProb{r}{d}\bigr)}$, as claimed.
\end{proofof}

\begin{proofof}{Proof of Theorem~\ref{thm:hardness}}
We reduce from the strongly NP-complete problem
\textsc{Product Partition}~\citep{ng2010product}: given positive
integers $a_1,\ldots,a_n$, determine whether there exists a partition
$(S_1,S_2)$ of $\{1,\ldots,n\}$ such that
\[
\prod_{j\in S_1}a_j=\prod_{j\in S_2}a_j.
\]

\paragraph{Reduction.}
Construct an instance with two identical riders and $n$ drivers,
one driver per integer. For each rider ${r}\in\{1,2\}$ and driver ${d\in\{1,\ldots,n\}}$,
set
\[
{\Weight{r}{d}=1,
\qquad
\AccProb{r}{d}=1-\frac{1}{a_d}.}
\]
By Lemma~\ref{lem:unitweights}, the FA and BA valuations coincide
and are monotone. We may therefore assume that all drivers are
assigned, so a feasible allocation is a partition $(S_1,S_2)$ of
$\DriverSet$.

\paragraph{Correctness.}
For any such allocation, define
\[
{A_r:=\prod_{d\in S_r}a_d,\qquad r\in\{1,2\},
\qquad
P:=\prod_{d=1}^n a_d,}
\]
where an empty product equals one. By
Lemma~\ref{lem:unitweights}, the total welfare under either
protocol is
\[
\Fi{1}{S_1}+\Fi{2}{S_2}
=
2-\frac{1}{A_1}-\frac{1}{A_2}.
\]
Since $A_1A_2=P$, the arithmetic--geometric mean inequality gives
\[
\Fi{1}{S_1}+\Fi{2}{S_2}
\le 2-\frac{2}{\sqrt{P}},
\]
with equality if and only if $A_1=A_2$.

Consequently, if the \textsc{Product Partition} instance is a
YES instance, every optimal allocation has $A_1=A_2$.
Conversely, an allocation satisfying $A_1=A_2$ directly gives a
valid product partition. Thus, an algorithm that computes an
optimal allocation allows us to decide \textsc{Product Partition}
by checking whether the two products are equal. These products
have polynomial binary encoding length, so this check takes
polynomial time.\end{proofof}

\subsection{Proof of \Cref{thm:NPhard}}
\begin{proofof}{Proof}
We reduce from \textsc{Subset Sum}. Consider an instance consisting of positive integers $a_1,\ldots,a_n$ and a target $B$. Let $A:=\sum_{i=1}^n a_i$, and assume without loss of generality that $1\le B\le A$. Define
\begin{align*} \varepsilon:=\frac{1}{10000A^3}. \end{align*}

Construct one distinguished driver $0$ with
\begin{align*} \pSingle{0}:=\frac12,\qquad \wSingle{0}:=1, \end{align*}
and, for each $i\in\{1,\ldots,n\}$, an item driver with
\begin{align*} \pSingle{i}:=\varepsilon a_i,\qquad \delta_i:=\frac{\varepsilon(2B-a_i)}{27},\qquad \wSingle{i}:=\frac13+\delta_i. \end{align*}
All probabilities and weights are rational numbers in $[0,1]$ with polynomial binary encoding length. Since driver $0$ has maximum weight, Lemma~\ref{lem:max-weight} implies that adding it to any notification set does not decrease its FA value. It therefore suffices to consider sets of the form $\{0\}\cup S$, where $S\subseteq\{1,\ldots,n\}$.

By multilinearity of the FA valuation in the acceptance probabilities, we may expand $F(\{0\}\cup S)$ in the item probabilities while keeping $\pSingle{0}=1/2$ fixed. The constant coefficient is $1/2$, the coefficient of $\pSingle{i}$ is
\begin{align*} \frac34\wSingle{i}-\frac14=\frac34\delta_i, \end{align*}
and the coefficient of $\pSingle{i}\pSingle{j}$ is
\begin{align*} \frac16-\frac{\wSingle{i}+\wSingle{j}}{3}=-\frac1{18}-\frac{\delta_i+\delta_j}{3}. \end{align*}
Consequently,
\begin{align*} F(\{0\}\cup S)=\frac12+\frac34\sum_{i\in S}\delta_i\pSingle{i}-\frac1{18}\sum_{\substack{i<j\\i,j\in S}}\pSingle{i}\pSingle{j}+R(S), \end{align*}
where $R(S)$ contains the pairwise terms involving $\delta_i+\delta_j$ and all terms of degree at least three.

Let $s:=\sum_{i\in S}a_i$. Substituting the definitions of $\pSingle{i}$ and $\delta_i$ gives
\begin{align*} \frac34\sum_{i\in S}\delta_i\pSingle{i}-\frac1{18}\sum_{\substack{i<j\\i,j\in S}}\pSingle{i}\pSingle{j} &=\frac{\varepsilon^2}{36}\sum_{i\in S}(2Ba_i-a_i^2)-\frac{\varepsilon^2}{18}\sum_{\substack{i<j\\i,j\in S}}a_ia_j \\ &=\frac{\varepsilon^2}{36}(2Bs-s^2)=\frac{\varepsilon^2}{36}\bigl(B^2-(s-B)^2\bigr). \end{align*}

We next bound the remainder. Since
\begin{align*} |\delta_i|\le\frac{2\varepsilon A}{27},\qquad P_S:=\sum_{i\in S}\pSingle{i}\le\varepsilon A, \end{align*}
the omitted pairwise terms have total absolute value at most
\begin{align*} \left|\frac13\sum_{\substack{i<j\\i,j\in S}}(\delta_i+\delta_j)\pSingle{i}\pSingle{j}\right|\le\frac{4\varepsilon A}{81}\sum_{\substack{i<j\\i,j\in S}}\pSingle{i}\pSingle{j}\le\frac{2\varepsilon A}{81}P_S^2\le \varepsilon^3A^3. \end{align*}
For every $k\ge3$, a degree-$k$ multilinear coefficient is an alternating sum of at most $2^k$ FA values. Since all weights lie in $[0,1]$, its absolute value is at most $2^k$. Hence, the total absolute contribution of all terms of degree at least three is bounded by
\begin{align*} \sum_{k=3}^{|S|}2^k\sum_{\substack{T\subseteq S\\|T|=k}}\prod_{i\in T}\pSingle{i}\le\sum_{k=3}^{\infty}\frac{(2P_S)^k}{k!}\le e(2P_S)^3<24\varepsilon^3A^3. \end{align*}
Combining these bounds and using the definition of $\varepsilon$, we obtain
\begin{align*} |R(S)|<25\varepsilon^3A^3=\frac{\varepsilon^2}{400}<\frac{\varepsilon^2}{144}. \end{align*}
Thus, for every $S\subseteq\{1,\ldots,n\}$,
\begin{align*} F(\{0\}\cup S)=\frac12+\frac{\varepsilon^2}{36}\bigl(B^2-(s-B)^2\bigr)+R(S),\qquad |R(S)|<\frac{\varepsilon^2}{144}. \end{align*}

Set the decision threshold to
\begin{align*} W:=\frac12+\frac{\varepsilon^2}{36}\left(B^2-\frac12\right). \end{align*}
If the \textsc{Subset Sum} instance is a YES instance, there exists $S$ with $s=B$, and therefore
\begin{align*} F(\{0\}\cup S)>\frac12+\frac{\varepsilon^2B^2}{36}-\frac{\varepsilon^2}{144}=W+\frac{\varepsilon^2}{144}>W. \end{align*}
Conversely, in a NO instance, $s$ and $B$ are integers with $s\ne B$, so $(s-B)^2\ge1$ for every $S$. Hence,
\begin{align*} F(\{0\}\cup S)<\frac12+\frac{\varepsilon^2(B^2-1)}{36}+\frac{\varepsilon^2}{144}=W-\frac{\varepsilon^2}{144}<W. \end{align*}
Sets omitting driver $0$ cannot exceed the threshold because adding driver $0$ does not decrease the FA value.

Thus, the \textsc{Subset Sum} instance is a YES instance if and only if the constructed single-rider FA instance has a notification set with value at least $W$. The construction is polynomial, proving that single-rider welfare maximization under FA is NP-hard.
\end{proofof}

\subsection{Proof of \Cref{lem:attenuation-integral}}
\begin{proofof}{Proof}
Fix $y\in[0,1]$ and define
\[
z(t):=1-tya(y)-e^{-ty},
\qquad
Z(t):=\int_0^t z(s)\,ds.
\]
The Taylor bound $e^{-ty}\leq1-ty+y^2t^2/2$ gives
\[
z(t)\geq y^2\left(\frac{t}{3}-\frac{t^2}{2}\right),
\]
and therefore
\[
Z(t)\geq\frac{y^2}{6}t^2(1-t)\geq0
\qquad\text{for all }t\in[0,1].
\]
For any nonnegative, nonincreasing, continuously differentiable function $h$ on $[0,1]$, integration by parts yields
\[
\int_0^1 h(t)z(t)\,dt
=
h(1)Z(1)-\int_0^1 h'(t)Z(t)\,dt
\geq0.
\]
Equivalently,
\[
\int_0^1 h(t)\bigl(1-tya(y)\bigr)\,dt
\geq
\int_0^1 h(t)e^{-ty}\,dt.
\]

Apply this inequality successively to the factors in the product, replacing $1-ty_\ell a(y_\ell)$ by $e^{-ty_\ell}$ one at a time. At each step, take $h$ to be the product of all remaining factors. Both the linear and exponential factors are nonnegative, nonincreasing, and continuously differentiable, so their product satisfies the required conditions. After all replacements, we obtain the desired inequality.
\end{proofof}

\subsection{Proof of Proposition~\ref{prop:BA-demand-DP}}
\label{apx:proof-DP-demand-oracle}
\begin{proofof}{Proof}
Fix rider $r$ and write $w_d \equiv w_{r,d}$ and $p_d \equiv p_{r,d}$ for brevity.
Recall that scores are normalized so that $w_d\in[0,1]$. Our algorithm follows four distinct steps:

\paragraph{Step 1: Pre-processing} First, we discard all drivers with $\lambda_d>1$ and assume that $\lambda_d\le 1$ for all $d$.
To see why this is without loss of generality, for any set $S$, we have $F_r(S)\le \max_{d\in S} w_d \le 1$ and $F_r(\emptyset)=0$.
Hence if $\lambda_d>1$, then for every $S\ni d$,
\[
F_r(S)-\sum_{k\in S}\lambda_k \;\le\; 1-\lambda_d \;<\; 0 \;\le\; F_r(\emptyset),
\]
so no optimal demand set ever includes such a driver. After this step, $\sum_{d\in\DriverSet}\lambda_d\le n$.

\paragraph{Step 2: Discretize prices.}
Let $K \coloneqq \epsilon/n$. For each driver $d$, define the rounded-down price
\[
\hat\lambda_d \;\coloneqq\; K\Bigl\lfloor \lambda_d/K \Bigr\rfloor,
\]
and for any set $S$ write $C(S)=\sum_{d\in S}\lambda_d$ and $\hat C(S)=\sum_{d\in S}\hat\lambda_d$.
Then for every $d$ we have $0\le \lambda_d-\hat\lambda_d < K$, and thus for every $S$,
\begin{equation}
\label{eq:BA-cost-rounding}
\hat C(S)\le C(S) < \hat C(S)+nK \;=\; \hat C(S)+\epsilon.
\end{equation}

\paragraph{Step 3: A DP for maximizing BA value under a discretized budget.}
Order drivers so that $w_1\ge w_2\ge \cdots \ge w_n$ (ties broken arbitrarily).
The key property of BA under this ordering is the following recursion:
if a set $S$ contains driver $t$ and $t$ is the highest-score driver in $S$, then
\[
F_r(S) \;=\; p_t w_t \;+\; (1-p_t)\,F_r(S\setminus\{t\}),
\]
because if $t$ accepts we obtain value $w_t$, and only if $t$ rejects does the outcome depend on
lower-score drivers. We now define a knapsack-style DP over the discretized prices. Let $B_{\max}\coloneqq \sum_{d\in\DriverSet}\hat\lambda_d$.
For each index $t\in\{1,\dots,n+1\}$ and each discretized budget
$B\in\{0,K,2K,\dots,B_{\max}\}$, let $V(t,B)$ be the maximum BA value achievable using only
drivers $\{t,t+1,\dots,n\}$ with total discretized cost at most $B$.
The DP recurrence (i.e., the Bellman update equation) is:
\[
V(t,B) \;=\;
\max\Bigl\{
V(t+1,B),\;
p_t w_t + (1-p_t)\,V(t+1,B-\hat\lambda_t)
\Bigr\},
\]
where the second term is available only when $B\ge \hat\lambda_t$.
We use boundary conditions $V(n+1,B)=0$ for all $B$, and $V(t,B)=-\infty$ for $B<0$.
We use standard back-tracking to recover an argmax set for each state $(t,B)$.

\paragraph{Step 4: Recover the best set for the discretized objective.}
For each budget $B$, the DP computes $\max\{F_r(S): \hat C(S)\le B\}=V(1,B)$.
We output the set $\hat S$ corresponding to a budget $\hat B$ that maximizes
\[
V(1,B)-B \quad \text{over } B\in\{0,K,2K,\dots,B_{\max}\}.
\]

We first analyze the correctness of the above algorithm by showing additive-$\epsilon$ guarantee for the original prices. Note that the choice $\hat{S}$ of the algorithm in Step~4 achieves:
\[
F_r(\hat S)-\hat C(\hat S) \;=\; \max_{S\subseteq\DriverSet}\bigl(F_r(S)-\hat C(S)\bigr),
\]
since for any set $S$ the choice $B=\hat C(S)$ is feasible in the budget optimization problem in Step~4 and yields an objective of $F_r(S)-\hat C(S)$, so:
$$
F_r(S)-\hat C(S)\leq V(1,B)-B \leq V(1,\hat B)-\hat B=F_r(\hat S)-\hat C(\hat S)~.
$$
Now, let $S^*$ maximize the true demand objective with prices $\{\lambda_d\}$. Using \eqref{eq:BA-cost-rounding} and optimality of $\hat S$ for the rounded objective:
\begin{align*}
F_r(\hat S)-C(\hat S)
&\ge F_r(\hat S)-\bigl(\hat C(\hat S)+\epsilon\bigr) \\
&= \bigl(F_r(\hat S)-\hat C(\hat S)\bigr)-\epsilon \\
&\ge \bigl(F_r(S^*)-\hat C(S^*)\bigr)-\epsilon \\
&\ge \bigl(F_r(S^*)-C(S^*)\bigr)-\epsilon.
\end{align*}
This proves the desired additive-$\epsilon$ approximation.

To analyze the running time, note that $B_{\max}\le \sum_d \lambda_d \le n$, and the budget grid has size
$B_{\max}/K = \mathcal{O}(n^2/\epsilon)$.
The DP has $\mathcal{O}(n\cdot n^2/\epsilon)$ states and $\mathcal{O}(1)$ work per state, hence runs in
$\mathcal{O}(n^3/\epsilon)$ time.
\end{proofof}

\end{document}